\documentclass[]{rsos}

\usepackage{caption}
\usepackage{capt-of}
\usepackage{subcaption}
\usepackage{enumitem}
\usepackage{placeins}

\usepackage{lineno}

\newcommand{\org}[1]{\textcolor{red}{}}
\newcommand{\new}[1]{\textcolor{black}{{#1}}}

\begin{document}

\title{A Computational Method for Sharp Interface Advection*}

\author{
Johan Roenby$^{1}$, Henrik Bredmose$^{2}$ and Hrvoje Jasak$^{3,4}$}

\address{$^{1}$DHI, Department of Ports \& Offshore Technology, Agern
All\'{e} 5, 2970 H\o rsholm, Denmark\\
$^{2}$Department of Wind Energy, Technical University of Denmark, 2800 Kgs. Lynbgy, Denmark\\
$^{3}$University of Zagreb, Faculty of Mechanical Engineering and Naval Architecture, Ivana Lu\v{c}i\'{c}a 5, Zagreb, Croatia\\
$^{4}$Wikki Ltd, 459 Southbank House, SE1 7SJ, London, United Kingdom}

\subject{Computational Fluid Dynamics, Multiphase flows, Numerical methods, Two phase flows, Free surface flows}


\corres{Johan Roenby\\
\email{jro@dhigroup.com}}

\begin{abstract}
We devise a numerical method for passive advection of a surface, such as the interface between two incompressible fluids, across a computational mesh. The method is called isoAdvector, and is developed for general meshes consisting of arbitrary polyhedral cells. The algorithm is based on the volume of fluid (VOF) idea of calculating the volume of one of the fluids transported across the mesh faces during a time step. The novelty of the isoAdvector concept consists in two parts: First, we exploit an isosurface concept for modelling the interface inside cells in a geometric surface reconstruction step. Second, from the reconstructed surface, we model the motion of the face-interface intersection line for a general polygonal face to obtain the time evolution \emph{within} a time step of the submerged face area. Integrating this submerged area over the time step leads to an accurate estimate for the total volume of fluid transported across the face. The method was tested on simple 2D and 3D interface advection problems both on structured and unstructured meshes. The results are very satisfactory both in terms of volume conservation, boundedness, surface sharpness, and efficiency. The isoAdvector method was implemented as an OpenFOAM\textregistered{} extension and is published as open source.
\end{abstract}

\begin{fmtext}
{* 1st revision submitted 8 June 2016}
\end{fmtext}


\maketitle

\section{Introduction}
In this paper we address the numerical challenge of advancing a surface moving in a prescribed velocity field. We will refer to this as the interface advection problem, since the surface often constitutes an interface e.g. between two fluids. Simple as the problem may appear, there is a large variety of problems in science, engineering, and industry where its solutions are far from trivial. Our motivation for addressing this problem is rooted in our usage of Computational Fluid Dynamics (CFD) as a practical engineering tool for calculating wave loads on coastal and marine structures. Whether it is an offshore wind turbine foundation, or an oil \& gas platform, accurate estimation of the peak loads from violent breaking waves is paramount for the correct dimensioning of the structure. In our view, CFD has a large unexploited potential to improve wave load estimates, and to reduce both cost and risks in the design phase of coastal and offshore structures. 

Due to the omnipresence of interfacial flows, the list of areas that may benefit from improved solution methods to the interface advection problem is almost endless. Some examples are bubble column reactors, oil-gas mixtures in pipelines, inkjet printing, automotive aquaplaning, ship manoeuvring, tank sloshing, dam breaks, metal casting processes, and hydraulic jumps.

During the past 40-50 years both Lagrangian and Eulerian strategies have been employed to develop a wide range of methods for advecting a sharp interface\cite{tryggvason_direct_2011}. We have been unable to find a recent review article dedicated to this vibrant research field. Today most CFD codes for practical engineering calculations use variants of the \emph{Volume-of-Fluid (VOF)} method for the interface advection step in their interfacial flow solvers. This includes current versions of ANSYS Fluent\textregistered, STAR-CCM+\textregistered, Gerris\cite{popinet_gerris:_2003}, OpenFOAM\textregistered \cite{OpenFoam,weller_tensorial_1998} and many others. In the VOF methodology the interface is implicitly represented via the \emph{volume fractions} of one of the fluids in computational cells. The advection is done by redistributing the content of this fluid between adjacent cells by moving it across the mesh faces. Since the first VOF methods appeared in literature\cite{hirt_volume_1981} a large variety of VOF schemes have been developed. They may be divided into two categories: \emph{Geometric} methods involving an explicit reconstruction of the interface from the volume fraction data, and \emph{algebraic} methods making no such attempt. Algebraic VOF schemes are typically much simpler to implement, more efficient, and are not restricted to structured meshes. They are, however, founded on much more heuristic considerations and are not as accurate as the geometric VOF schemes\cite{deshpande_evaluating_2012}. Geometric VOF schemes, on the other hand, involve complex geometric operations making their implementation cumbersome and their execution slow. Geometric VOF methods for unstructured meshes is an active area of research\cite{ahn_multi-material_2007, hernandez_new_2008, lopez_new_2008, ivey_conservative_2012, maric_vofoam_2013, xie_efficient_2014}. 

Our ambition in the development of the isoAdvector algorithm is to develop a VOF based interface advection method that works on arbitrary meshes, retains the accuracy of the geometric schemes by explicitly approximating the interface, and yet keeps the geometric operations at a minimum in order to obtain acceptable calculation times. An efficient VOF scheme yielding accurate results even on automatically generated unstructured meshes of  complex geometries has a huge potential for speeding up the simulation process and making CFD an integrated part of design processes involving interfacial flows.

In the remainder of this section we give an introduction to the interface advection problem and its formulation in the VOF framework. In Section \ref{sec:advMeth} we present the new ideas and concepts of the isoAdvector method and give an overview of the steps involved in the numerical procedure. The implementation details and considerations involved in each step are described at length in Section~\ref{sec:implementation}. In Section~\ref{sec:results} we demonstrate the performance of the new method with a series of simple test cases. Finally, in Section~\ref{sec:conclusion} we summarize our findings.

\subsection{VOF formulation of the interface advection problem}
We consider a computational domain $\mathcal D\in \mathbb{R}^3$ in which a surface $\mathcal S$ is embedded. The surface may consist of any number of closed surfaces and may also extend to the boundaries of the domain. We will think of the surface, $\mathcal S$, as the interface between two incompressible, immiscible fluids denoted by A and B, and occupying the two closed regions, $\mathcal A$ and $\mathcal B$, satisfying $\mathcal A\cap\mathcal B = \mathcal S$ and $\mathcal A\cup\mathcal B = \mathcal D$.

The fluid particles are assumed to be passively advected in a continuous, solenoidal velocity field, $\mathbf u(\mathbf x,t)$, which is defined in the whole domain, $\mathcal D$. In practical engineering applications involving incompressible two-phase flows the time evolution of the velocity field is governed by the Navier-Stokes equations for $\mathbf u$ coupled with a Poisson equation for the pressure, $p$. This system of equations must be solved simultaneously with the interface advection problem. In this work, we focus entirely on the interface advection problem, thus assuming $\mathbf u(\mathbf x,t)$ to be known in advance for all points, $\mathbf x\in\mathcal D$, and all times, $t$. 

We will now represent the surface $\mathcal S(t)$ in terms of a density field, $\rho(\mathbf x,t)$, which takes one constant value, $\rho_A$, everywhere in $\mathcal A$ and another constant value, $\rho_B$, everywhere in $\mathcal B$. The density field thus has a discontinuity at the interface $\mathcal S$.\footnote{On the surface $\mathcal S$ one could set $\rho = \frac12(\rho_A+\rho_B)$ for $\rho$ to be defined everywhere. However, since $\mathcal S$ has zero volume, the value of $\rho$ on $\mathcal S$ is immaterial.} The evolution of the surface is then determined by the integral form of the continuity equation,
\begin{equation}\label{eq:intgralContEqn}
	\frac{d \ }{dt}\int_{\mathcal V} \rho(\mathbf x,t) dV = - \int_{\partial \mathcal V} \rho(\mathbf x,t) \mathbf u(\mathbf x,t)\cdot d\mathbf S,
\end{equation}
where $\mathcal V\in\mathcal D$ is an arbitrary \new{stationary} volume, $\partial \mathcal V$ is its boundary, and $d\mathbf S$ is the differential area vector pointing out of the volume. \new{In words, this mass conservation equation says, that the instantaneous rate of change of the total mass enclosed in the volume is given by the instantaneous flux of mass through its boundary.}

In the pure advection problem with a predetermined velocity field the specific values of the fluid densities, $\rho_A$ and $\rho_B$ are immaterial, that is, the solution does not depend on them. To remove these insignificant parameters from the problem, we define the indicator field,
\begin{equation}\label{eq:indicatorField}
	H(\mathbf x,t) \equiv \frac{\rho(\mathbf x,t)-\rho_B}{\rho_A-\rho_B},
\end{equation}
such that $H = 1$ for all $\mathbf x\in\mathcal A(t)$, and $H = 0$ for all $\mathbf x\in\mathcal B(t)$. 

We now discretise the computational domain, $\mathcal D$, by conceptually dividing it into a large number of control volumes, or \emph{cells}, $\mathcal C_i$, for $i=1,...,N_C$. If two cells $i$ and $j$ are adjacent, their shared boundary, $\partial \mathcal C_i\cap \partial \mathcal C_j$ is called an \emph{internal face}. If cell $i$ touches the domain boundary, the shared surface $\partial \mathcal C_i\cap \partial \mathcal D$, will consist of one or more \emph{boundary faces}. All faces are labelled with integers, $j = 1,...,N_F$, and the surface of face $j$ is denoted $\mathcal F_j$. Thus the boundary of the cell $i$, may be represented by a list, $B_i$, of all the labels of faces belonging to its boundary, $\partial \mathcal C_i$. 

With these mesh definitions in place, we can now substitute \eqref{eq:indicatorField} into \eqref{eq:intgralContEqn} with cell $i$ as the volume of integration,
\begin{equation}\label{eq:CVintgratedContEqn}
	\frac{d \ }{dt}\int_{\mathcal C_i} H(\mathbf x,t) dV = - \sum_{j\in B_i}s_{ij}\int_{\mathcal F_j} H(\mathbf x,t) \mathbf u(\mathbf x,t)\cdot d\mathbf S.
\end{equation}
Because face $j$ has its own orientation determining the direction of $d\mathbf S$, we have introduced the auxiliary factor $s_{ij} = +1$ or $-1$, such that $s_{ij} d\mathbf S$ points out of cell $i$ for face $j$.

The natural next step is to define the volume fraction of fluid A in cell $i$, 
\begin{equation}\label{eq:volFracDef}
	\alpha_i(t) \equiv \frac1{V_i}\int_{\mathcal C_i} H(\mathbf x,t) dV,
\end{equation}
where $V_i$ is the volume of cell $i$. Substituting \eqref{eq:volFracDef} into \eqref{eq:CVintgratedContEqn}, and formally integrating \eqref{eq:CVintgratedContEqn} from time $t$ to time $t+\Delta t$, we obtain the following equation for the updated volume fraction of cell $i$,
\begin{equation}\label{eq:newAlpha}
	\alpha_i(t+\Delta t) = \alpha_i(t) - \frac1{V_i}\sum_{j\in B_i}s_{ij}\int_{t}^{t+\Delta t}\int_{\mathcal F_j} H(\mathbf x,\tau) \mathbf u(\mathbf x,\tau)\cdot d\mathbf S d\tau.
\end{equation}
We stress that this equation is exact with no numerical approximations introduced yet. It is the fundamental equation from which \org{we will derive our new}\new{one must derive any consistent} interface advection method. The time integral on the right hand side is the total volume of fluid A transported across face $j$ during the time interval from time $t$ to $t+\Delta t$. It is the fundamental quantity that we must estimate in order to advance $\alpha_i$, and hence implicitly the surface $\mathcal S$, in time. We will denote this quantity
\begin{equation}\label{eq:DeltaVDef}
	\Delta V_j(t,\Delta t) \equiv \int_{t}^{t+\Delta t}\int_{\mathcal F_j} H(\mathbf x,\tau) \mathbf u(\mathbf x,\tau)\cdot d\mathbf S d\tau.
\end{equation}
\org{Our}\new{The} fundamental equation \eqref{eq:newAlpha} can then be formulated as
\begin{equation}\label{eq:finalAlpha}
	\alpha_i(t+\Delta t) = \alpha_i(t) - \frac1{V_i}\sum_{j\in B_i}s_{ij}\Delta V_j(t,\Delta t).
\end{equation}

Before we move on to present the basic ideas of the isoAdvector method, we will need to consider how the velocity field is represented. In the finite volume treatment of the fluid equations of motion the natural representation of the velocity field is in terms of cell averaged values,
\begin{equation}
	\mathbf u_i(t) \equiv \frac1{V_i}\int_{\mathcal C_i} \mathbf u(\mathbf x, t) dV.
\end{equation}
Since the convective terms in the governing fluid equations give the transport of mass, momentum, etc. across cell faces, another important velocity field representation are the volumetric fluxes across mesh faces,
\begin{equation}\label{eq:fluxdef}
	\phi_j(t) \equiv \int_{\mathcal F_j} \mathbf u(\mathbf x,t)\cdot d\mathbf S.
\end{equation}
The question we will try to answer in the following can now be formulated as follows:
\\
\\
\emph{How do we most accurately and efficiently exploit the available information at time $t$, i.e. the volume fractions, $\alpha_i$, and the velocity data, $\mathbf u_i$ and $\phi_j$, to estimate the fluid A volume transport, $\Delta V_j(t,\Delta t)$, across a face during the time interval $[t,t+\Delta t]$?}

\section{The isoAdvector concept}\label{sec:advMeth}
We will now present the general ideas behind the isoAdvector method, starting with the interface representation using isosurfaces, then introducing the concept of a \emph{face-interface intersection line} moving across a face, and finally giving an overview of the steps involved in the numerical procedure. For the sake of clarity, we focus on ideas in this section, and postpone the detailed description of the implementation to Section \ref{sec:implementation}. \new{For full implementation details, the reader is referred to the source code provided with this article\cite{isoAdvector}}.

\subsection{The interface reconstruction step}\label{ssec:IntRecon}
The integral in \eqref{eq:DeltaVDef} is highly dependent on the local distribution of fluid A and B inside cell $i$ and inside its neighbour cells from which it receives fluid during the time step. However, the volume fractions hold no information about the distribution of the two fluids inside the cells. We must therefore come up with a subgrid model for this ``intracellular'' distribution from the given volume fraction data. If the volume fraction data is ``sharp'', only cells very close to the interface will have volume fractions significantly different from $0$ and $1$. Then, if cell $i$ is on the interface, its neighbours in one direction will mainly contain fluid A, while its neighbour cells in the opposite direction will mainly contain fluid B. In words, we want our subgrid model to \org{``detect'' this}\new{capture this local distribution information}, and place the fluid A content of cell $i$ close to the neighbours containing fluid A (which is equivalent to its fluid B content being placed near the neighbours containing fluid B). The implicit assumption made in this model is that the interface is sufficiently well resolved by the mesh such that \org{the two fluids occur in lumps somewhat larger than the cell size}\new{the local radius of curvature is larger than the cell size}. Whenever this is satisfied an isosurface calculation will provide \org{exactly}\new{a good estimate of the} the required \org{``lump''}\new{local fluid distribution} information. 

\begin{figure}[!tb]
\begin{center}
    \begin{subfigure}[t]{0.48\textwidth}
	\includegraphics[width=\linewidth]{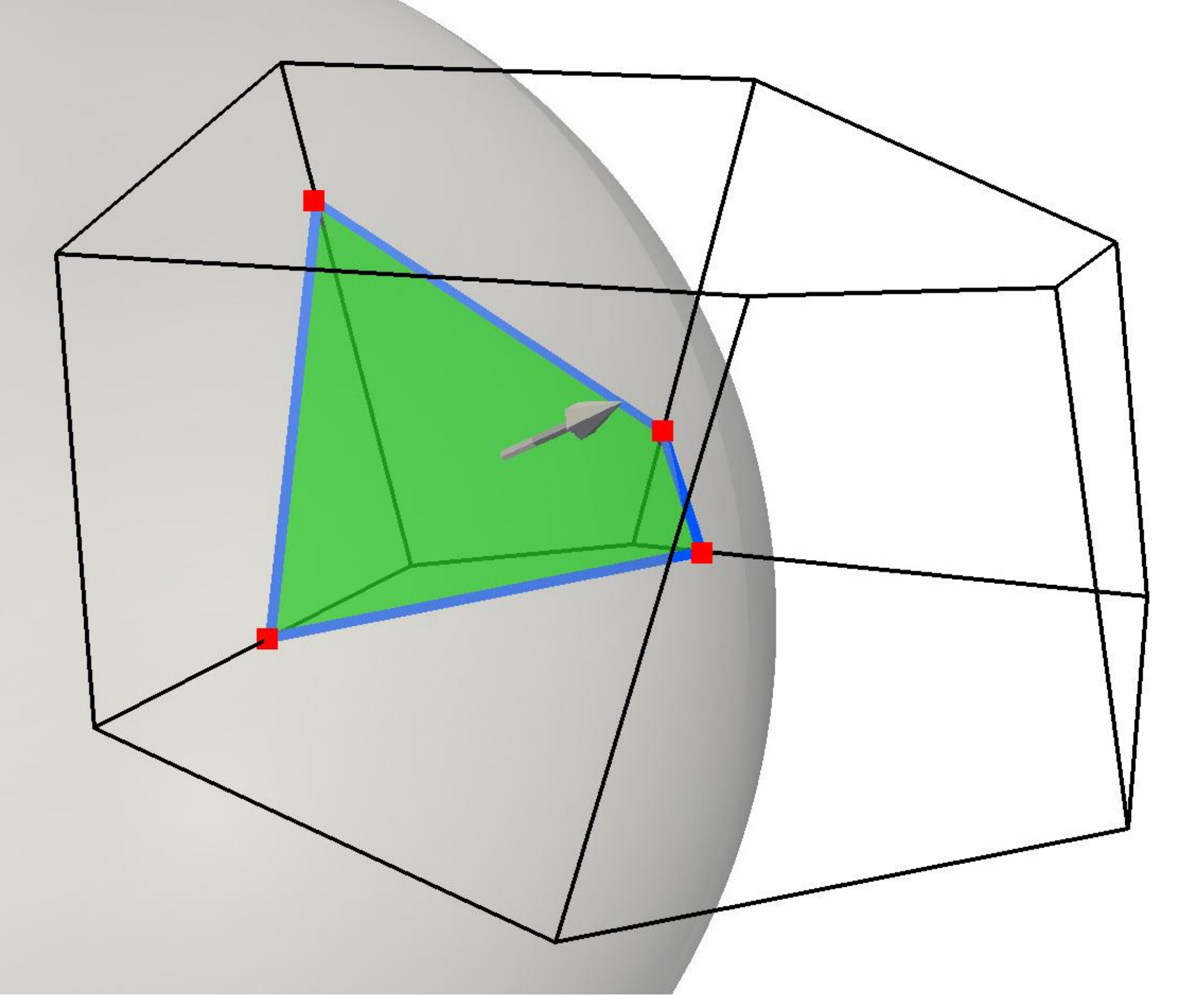}
	\caption{\quad}
    \label{fig:isoface}
    \end{subfigure}
    \quad
    \begin{subfigure}[t]{0.48\textwidth}
	\includegraphics[width=\linewidth]{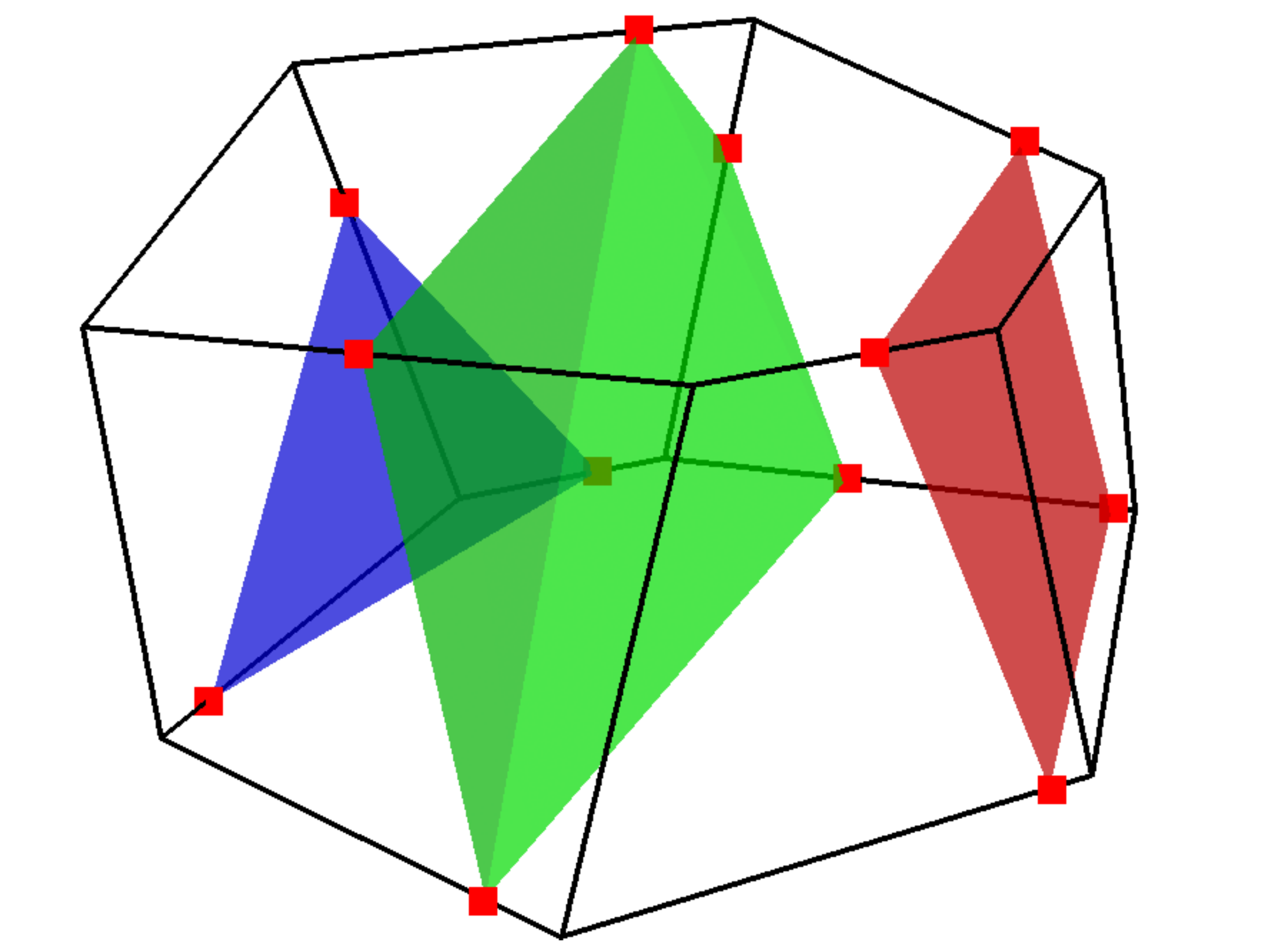}
	\caption{\quad}
    \label{fig:isofaceMotion}
    \end{subfigure}
	\caption{(a) A spherical surface cutting a polyhedral cell. Red dots are the edge cutting points. Blue lines are the face-interface intersection lines. Green patch is the isoface. (b) The isoface motion is estimated from surrounding velocity data and the isoface is propagated. Isoface at three different times within a time step are shown.}\label{fig:isofaceIllustration}
\end{center}
\end{figure}

The idea of using an isosurface numerically calculated from the volume fractions to represent the interface is inspired by our use of visualisation software, such as ParaView\textregistered\cite{ParaView}, for visualising surfaces. Numerically calculated isosurfaces are topologically consistent continuous surfaces and straightforward to calculate on arbitrary polyhedral meshes. The numerical representation of an isosurface in a polyhedral cell is a list of the points, where the isosurface cuts the cell edges. See red points in Fig.~\ref{fig:isoface} for an illustration. This list of points represents a face, which cuts the cell into two polyhedral subcells, with one completely immersed in fluid A, and the other completely immersed in fluid B. We will call such a face an \emph{isoface}. See the green patch in Fig.~\ref{fig:isoface} for an example. \new{We note that if an isoface has more than three vertices, it will generally not be exactly planar.}
When calculating an isosurface from the volume fraction data, we have the freedom of choosing an isovalue between 0 and 1. Which isovalue should we choose? For surface visualisation from volume fraction data, we usually plot the 0.5-isosurface. This, however, is not a good choice for the surface reconstruction step in an interface advection algorithm, because the isoface in cell $i$ with isovalue 0.5 does not in general cut it into two subcells of the volumetric proportions dictated by the volume fraction, $\alpha_i$. It may for instance occur, that the cell has $\alpha_i = 0.8$, and yet is not even cut by the $0.5$-isosurface. Hence, a surface reconstruction model based on the $0.5$-isosurface would say nothing about how the $80\%$ fluid A and $20\%$ fluid B is distributed inside this cell. There will, however, exist an isoface with another isovalue, which will cut the cell into subcells of the correct volumetric proportions. An important component of our proposed scheme is an efficient method for finding this isovalue for a given surface cell (see Section \ref{sec:implementation} for details). 

We note that with the use of different isovalues in different surface cells, the union of isofaces is no longer a continuous surface, as it would be if the same isovalue was used in adjacent cells. \new{This is an unavoidable price we must pay to ensure that isofaces cut surface cells into subcells having the correct volumetric proportions.}

\subsection{The advection step}
Most Navier-Stokes solvers for interface flows use a segregated solution approach, in which the coupled system of equations governing the flow are solved in sequence within a time step. This means, that at the point where the interface is to be advected from time $t$ to time $t+\Delta t$, we only have information about the velocity field up to time $t$. But as seen in the integrand in \eqref{eq:newAlpha}, calculation of the updated $\alpha_i$ requires information about the velocity field on the interval $[t,t+\Delta t]$. We must therefore estimate the evolution of the velocity field during the time step. The simplest such estimate is to regard the velocity field as constant (in time) during the whole time step. With this assumption we can write $\mathbf u(\mathbf x,\tau) \approx \mathbf u(\mathbf x,t)$ in \eqref{eq:DeltaVDef}. Another assumption we will make in \eqref{eq:DeltaVDef}, is that $\mathbf u$ on the face $\mathcal F_j$ dotted with the differential face normal vector, $d\mathbf S$, can be approximated in terms of the volumetric face flux, $\phi_j$ (defined in \eqref{eq:fluxdef}), as follows,
\begin{equation}
	\mathbf u(\mathbf x,t)\cdot d \mathbf S \approx \frac{\phi_j(t)}{|\mathbf S_j|} dS \textrm{ for } \mathbf x\in\mathcal F_j, 
\end{equation}
where $dS \equiv d|\mathbf S|$, and the face normal,
\begin{equation}
	\mathbf S_j \equiv \int_{\mathcal F_j} d\mathbf S.
\end{equation}
Substituting this into \eqref{eq:DeltaVDef}, we obtain
\begin{equation}\label{eq:DeltaVApprox1}
	\Delta V_j(t,\Delta t) \approx \frac{\phi_j(t)}{|\mathbf S_j|}\int_{t}^{t+\Delta t}\int_{\mathcal F_j} H(\mathbf x,\tau) dS d\tau.
\end{equation}
The remaining surface integral in \eqref{eq:DeltaVApprox1} is then simply the instantaneous area of face $j$ submerged in fluid A, which will be denoted
\begin{equation}\label{eq:A_j(t)Def}
	A_j(\tau) \equiv \int_{\mathcal F_j} H(\mathbf x,\tau) dS = \int_{\mathcal F_j \cap \mathcal A(\tau)}  dS.
\end{equation}
Using this definition, we may now write \eqref{eq:DeltaVApprox1} as
\begin{equation}\label{eq:DeltaVApprox2}
	\Delta V_j(t,\Delta t) \approx \frac{\phi_j(t)}{|\mathbf S_j|}\int_{t}^{t+\Delta t}A_j(\tau)d\tau.
\end{equation}
An important point is that in the special case, where the velocity field is constant both in space and time, \eqref{eq:DeltaVApprox2} is exact. Thus, if the cells become sufficiently small compared to the gradients of the velocity field, and the time steps become sufficiently small compared to the temporal variations in the velocity field, the error committed in the above approximation becomes negligible.

As is seen from \eqref{eq:DeltaVApprox2}, the challenge in constructing a VOF scheme is to estimate the  time evolution \emph{within} a time step of the submerged (in fluid A) area of a face, and then integrate this area in time. The time scale on which $A_j(\tau)$ changes is not dictated by the time scales of the flow, but by a complicated combination of the relative orientations of the face and interface, the direction of motion of the interface, and the shape of the \new{specific polygonal} face. As an example, consider an interface approaching a face to which it is parallel. In this case $A_j(\tau)$ will be a discontinuous function of $\tau$. This in turn makes $\Delta V_j(t,\Delta t)$ non-differentiable with respect to $\Delta t$. The discontinuous and non-differentiable nature of these quantities is what makes the interface advection problem so difficult to attack with the traditional weaponry of numerical analysis, which assumes the existence of a Taylor expansion of the sought solution.

In the isoAdvector advection step, when we calculate $A_j(\tau)$ for face $j$, our starting point is the isoface in the cell \emph{upwind} of face $j$ at time $t$\new{, because this is the cell from which the face receives fluid during the time step}. The motion of this isoface \emph{within} the time step $[t,t+\Delta t]$ may be approximated by using the velocity data in the surrounding cells. Fig.~\ref{fig:isofaceMotion} shows an example of how the isoface may appear at three times during the time step. \new{For details on our approximation of the isoface motion, the reader is referred to Section~\ref{ssec:EstIsofaceMotion}.}

Knowing the isoface position and orientation inside cell $i$ at any time within $[t,t+\Delta t]$, we also know for its downwind face $j$ the face-interface intersection line (see blue lines in Fig.~\ref{fig:isoface}) at any time during the time interval. With this information, the time integral in \eqref{eq:DeltaVApprox2} can be calculated \new{analytically}, to finally obtain our estimate of the total volume of fluid A transported across face $j$ during the time interval $[t,t+\Delta t]$. 

We stress, that the fluid A transport across a face is only calculated once for each face, and that for internal faces, this value is used to update the volume fractions of \emph{both} of the two cells sharing the face. This \org{is what }guarantees local \new{and global} conservation of each of the two fluids A and B.

\subsection{Algorithm overview}\label{ssec:algorithm}
We here give an overview of the steps taken in the isoAdvector algorithm to advance the volume fractions from time $t$ to time $t+\Delta t$:

\begin{description}
   \item[Step 1] For each face $j$ initialise $\Delta V_j$ with the upwind cell volume fraction, $\Delta V_j = \alpha_{\textrm{upwind}(j)}\phi_j\Delta t$.
   \item[Step 2] Find all \emph{surface cells}, i.e. cells with $\epsilon < \alpha_i(t) < 1-\epsilon$, where $\epsilon$ is a user specified tolerance (we typically use $10^{-8}$).
   \item[Step 3] For each surface cell $i$, do the following:

   \begin{description}
   		\item[3.1] Find its isoface, i.e. the isosurface inside the cell with isovalue such that it cuts the cell into the correct volumetric fractions, $\alpha_i(t)$ and $1-\alpha_i(t)$ (Details in Section~\ref{ssec:calcIsoface}).\label{item:findIsoface}
   		\item[3.2] Use the velocity field data to estimate the isoface motion during the time interval $[t,t+\Delta t]$ (Details in Section~\ref{ssec:EstIsofaceMotion}).\label{item:findIsofaceMotion}
   		\item[3.3] For each \emph{downwind} face $j$ of surface cell $i$, use the isoface and its motion to calculate the face-interface intersection line during the time interval $[t,t+\Delta t]$ (Details in Section~\ref{ssec:EvoOfFiil}).\label{item:findFaceIsofaceIntersection}
   		\item[3.4] For each \emph{downwind} face $j$ of surface cell $i$, use the motion of its face-interface intersection line to calculate $\Delta V_j(t,\Delta t)$ from the time integral in \eqref{eq:DeltaVApprox2} (Details in Section \ref{ssec:timeIntSubArea}).\label{item:integrateFaceIsofaceIntersection}
	\end{description}

	\item[Step 4] For each cell calculate $\alpha_i(t+\Delta t)$ by inserting the $\Delta V_j$'s of its faces in \eqref{eq:finalAlpha}.\label{item:updateVOF}
	\item[Step 5] For cells with $\alpha_i(t+\Delta t) < 0$ or $\alpha_i(t+\Delta t) > 1$ adjust the $\Delta V_j$'s of its faces using a redistribution procedure and recalculate $\alpha_i(t+\Delta t)$ by inserting corrected $\Delta V_j$'s in \eqref{eq:finalAlpha}. This step also includes an optional \new{subsequent} clipping of any \new{cell values} $\alpha_i < 0$ or $\alpha_i > 1$ to ensure strict boundedness before proceeding to next time step (Details in Section~\ref{ssec:bounding}).\label{item:boundVOF}
\end{description}

\section{Implementation details}\label{sec:implementation}

In this section, we provide the implementation details of the procedure outlined in Section \ref{ssec:algorithm}. We first note, that the time step size may vary between time steps. The user can specify a target \emph{interface} Courant number, Co, based on which the time step size is set at the beginning of each time step, to ensure that Co is not exceeded in any surface cells. \new{The interface Courant number only concerns the velocity of the interface normal to itself in surface cells. }

Step 1 in Section \ref{ssec:algorithm}, where we initialise $\Delta V_j$ with upwind values, and Step 2, where we find all surface cells with $\epsilon < \alpha_i < 1-\epsilon$, need no further explanation. We will therefore jump to step 3, which contains the actual calculation of the volume transport across faces.

\subsection{Calculating the initial isoface in a surface cell}\label{ssec:calcIsoface}

The first step in calculating the isosurface is to interpolate the volume fractions to the mesh points. The value in a mesh point will in general be a linear combination of the volume fractions in the cells sharing the mesh point. We have chosen to use inverse point-to-cell-centre interpolation, but other options, such as cell volume weighting, are also possible. 

Let us temporarily denote the $N$ vertices of cell $i$ by $\mathbf X_1,...,\mathbf X_N$ and the corresponding interpolated volume fractions by $f_1,...,f_N$. The cell edges are straight lines between pairs of points in the vertex list. To construct the $f$-isoface for cell $i$, we go through all cell edges and cut them by linear interpolation of the edge vertex values: If the edge $(\mathbf X_k, \mathbf X_l)$ has values $f_k < f$ and $f < f_l$, the edge is cut at the point
\begin{equation}
	\mathbf x_{\textrm{cut}} = \mathbf X_k + \frac{f-f_k}{f_l-f_k}(\mathbf X_l-\mathbf X_k).
\end{equation}
Once all such edge cutting points have been found for cell $i$, they can be connected across faces to form the face-interface intersection lines, which can again be connected to form the isoface inside the cell (see Fig.~\ref{fig:isoface}). The isoface splits cell $i$ into a polyhedral cell, $\mathcal A_i(f)$, entirely in fluid A, and another cell, $\mathcal B_i(f)$, entirely in fluid B. We can calculate the volume of $\mathcal A_i(f)$ relative to the cell volume,
\begin{equation} 
	\tilde\alpha(f) = \frac{\textrm{vol}(\mathcal A_i(f))}{V_i}.
\end{equation}
This will vary monotonically and continuously from $0$ to $1$, as the isovalue $f$ varies from the maximum vertex value, max$(f_1,...,f_N)$, to the minimum vertex value, min$(f_1,...,f_N)$. As argued in Section \ref{ssec:IntRecon}, the correct isovalue to use is the one recovering the cell volume fraction. That is, we should find $f^*$ such that $\tilde\alpha(f^*) = \alpha_i$. In the current implementation $f^*$ is found by \org{simple bisection root finding}\new{the following procedure: First, we geometrically calculate $\tilde\alpha(f)$ for the vertex values, $f_1,...,f_N$, finding the two closest values, say, $f_k$ and $f_l$, such that $f^*\in[f_k, f_l]$. Between these values, we know that $\tilde\alpha(f)$ increases monotonically like a cubic polynomial. Thus, evaluating $\tilde\alpha(f)$ geometrically at two more points in between $f_k$ and $f_l$, we have 4 equations for the 4 polynomial coefficients. The resulting linear 4$\times$4 matrix system we solve using LU decomposition. With a polynomial expression for $\tilde\alpha(f)$ at hand, we can use Newton's root finding method to efficiently find $f^*$ such $|\tilde \alpha(f^*)-\alpha_i|<\epsilon$, where $\epsilon$ is a user specified tolerance, typically set to $\epsilon = 10^{-8}$. In rare cases the LU solution does not give useful coefficients because the 4$\times$4 matrix is ill-conditioned, so the method does not converge. In these cases, we use Newton's root finding method with direct geometric evaluation of $\tilde\alpha(f)$ instead of the much cheaper polynomial evaluation.}\org{\footnote{\org{Because $\tilde \alpha(f)$ is only piecewise smooth it is not a good idea to use gradient based root finding methods to try to improve convergence.}} The root finding stops when $|\tilde \alpha(f^*)-\alpha_i|<\epsilon$, where $\epsilon$ is a user specified tolerance. We typically set $\epsilon = 10^{-8}$.}\footnote{\org{A potentially more efficient way of finding $f^*$ is the following: Suppose $\tilde f_1<...<\tilde f_K$ is the sorted list of unique vertex values. We can use binary search to quickly find the interval $[\tilde f_l,  \tilde f_{l+1}]$ in which $f^*$ must be. On this interval the function $\tilde \alpha(f)$ varies as a cubic polynomial. Thus, if we evaluate $\tilde \alpha(f)$ at $f_l$ and $f_{l+1}$, and at two points in between, we have four equations for the four coefficients in the cubic polynomial. Solving these equations, we obtain the exact functional expression for $\tilde \alpha(f)$ on the relevant $f$-interval. This allows us to efficiently find $f^*$, either by analytic evaluation of the polynomial roots, or by a numerical root finding algorithm with a specified tolerance.}}

We note, that, due of to the cell-to-vertex interpolation, the effective stencil contributing to the isoface inside a surface cell consists of the cell itself with all its point neighbours, that is, all surrounding cells with which it shares a vertex. 

\subsection{Estimating the isoface motion during a time step}\label{ssec:EstIsofaceMotion}
We first calculate the geometric face centre, $\mathbf x_S$, and the unit normal vector, $\hat{\mathbf n}_S$, of the isoface (see Fig.~\ref{fig:isoface}). The procedure for doing this is the same as for a \org{regular}\new{any other} mesh face in OpenFOAM\textregistered{}: The average point between the $N$ vertex points of the N-gonal face is calculated, and the face is decomposed into $N$ triangles all sharing this average point as their common top point. The face centre, $\mathbf x_S$, is then calculated as the area weighted average of the geometric centres of these $N$ triangles. Likewise, the face normal vector, $\mathbf n_S$, is calculated as the area weighted average of the $N$ triangle area vectors. 

The next step is to interpolate the velocity data, $\mathbf u_i(t)$, to $\mathbf x_S$. This is done by first decomposing the cell into tetrahedra all sharing the cell centre as their common top point. Then we find the tetrahedron containing $\mathbf x_S$, and interpolate the velocity field into its vertices. Finally we interpolate linearly from the tetrahedral vertices to obtain the velocity vector $\mathbf U_S$ at $\mathbf x_S$. \new{We note that for stationary meshes the weightings in this interpolation procedure only need to be calculated and stored once at the beginning of a simulation.}

The next step is to dot $\mathbf U_S$ with the isoface normal, $\hat{\mathbf n}_S$, to obtain the isoface motion normal to itself, $U_S \equiv \mathbf U_S\cdot \hat{\mathbf n}_S$. We will make the convention that $\hat{\mathbf n}_S$ is directed from fluid A into fluid B. Thus, positive $U_S$ means that the cell is filling up with fluid A, while negative $U_S$ means that it is filling up with fluid B. In the current implementation, we regard \new{the} $U_S$ \new{of an isoface} as constant during the whole time step. Possible improvements could be 1) using velocity data from previous time steps to estimate the isoface acceleration during the time step and 2) calculating the velocity gradient from surrounding cell velocity data to approximate the isoface rotation around its two tangential axes during the time step. \new{Work along these lines is left for future development.}

\subsection{Evolution of the face-interface intersection line}\label{ssec:EvoOfFiil}
\begin{figure}[!tb]
\begin{center}
    \begin{subfigure}[b]{0.49\textwidth}
		\includegraphics[width=\linewidth]{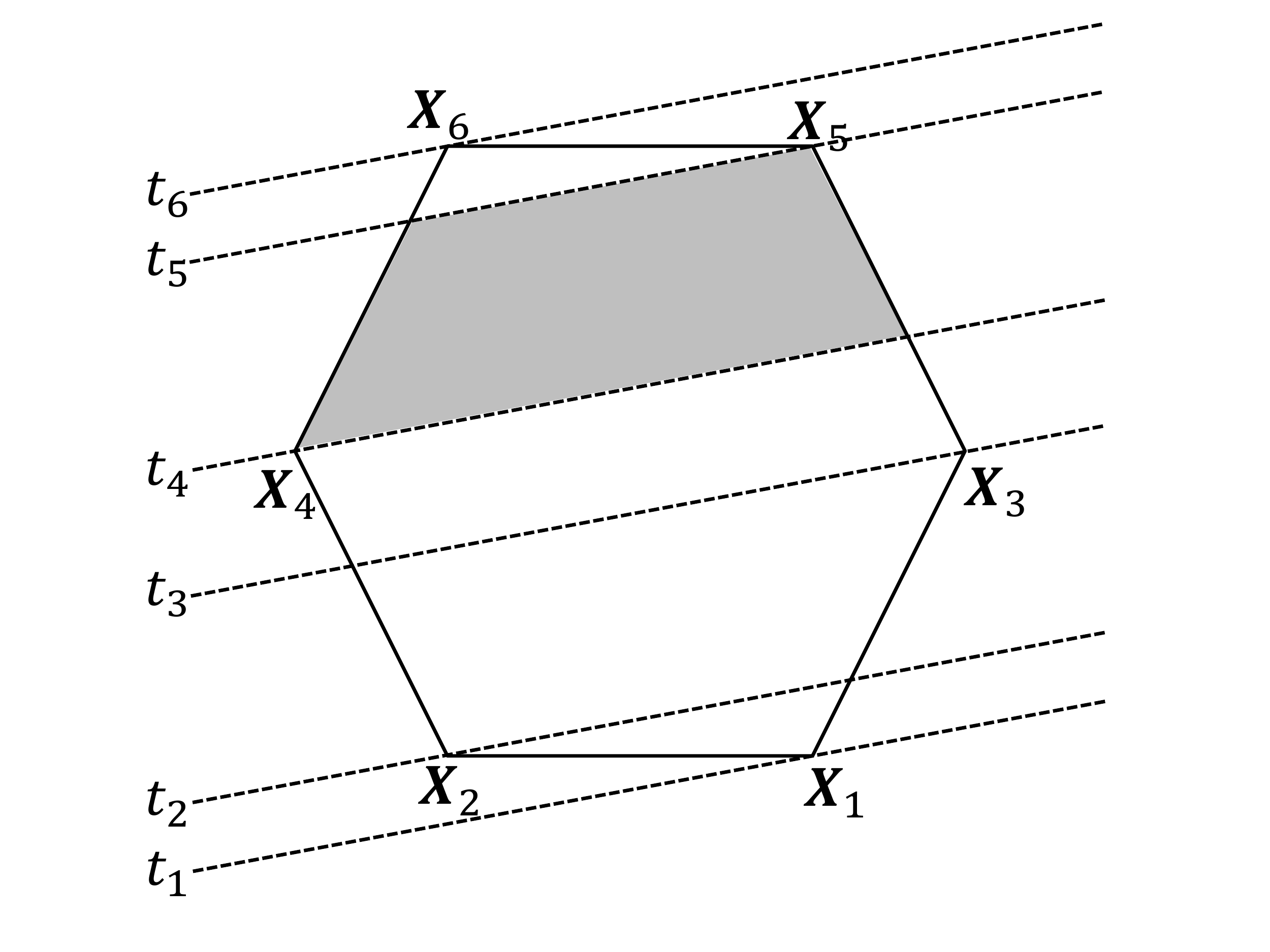}
		\caption{}
		\label{subfig:sweptArea}
    \end{subfigure}
    \begin{subfigure}[b]{0.49\textwidth}
		\includegraphics[width=\linewidth]{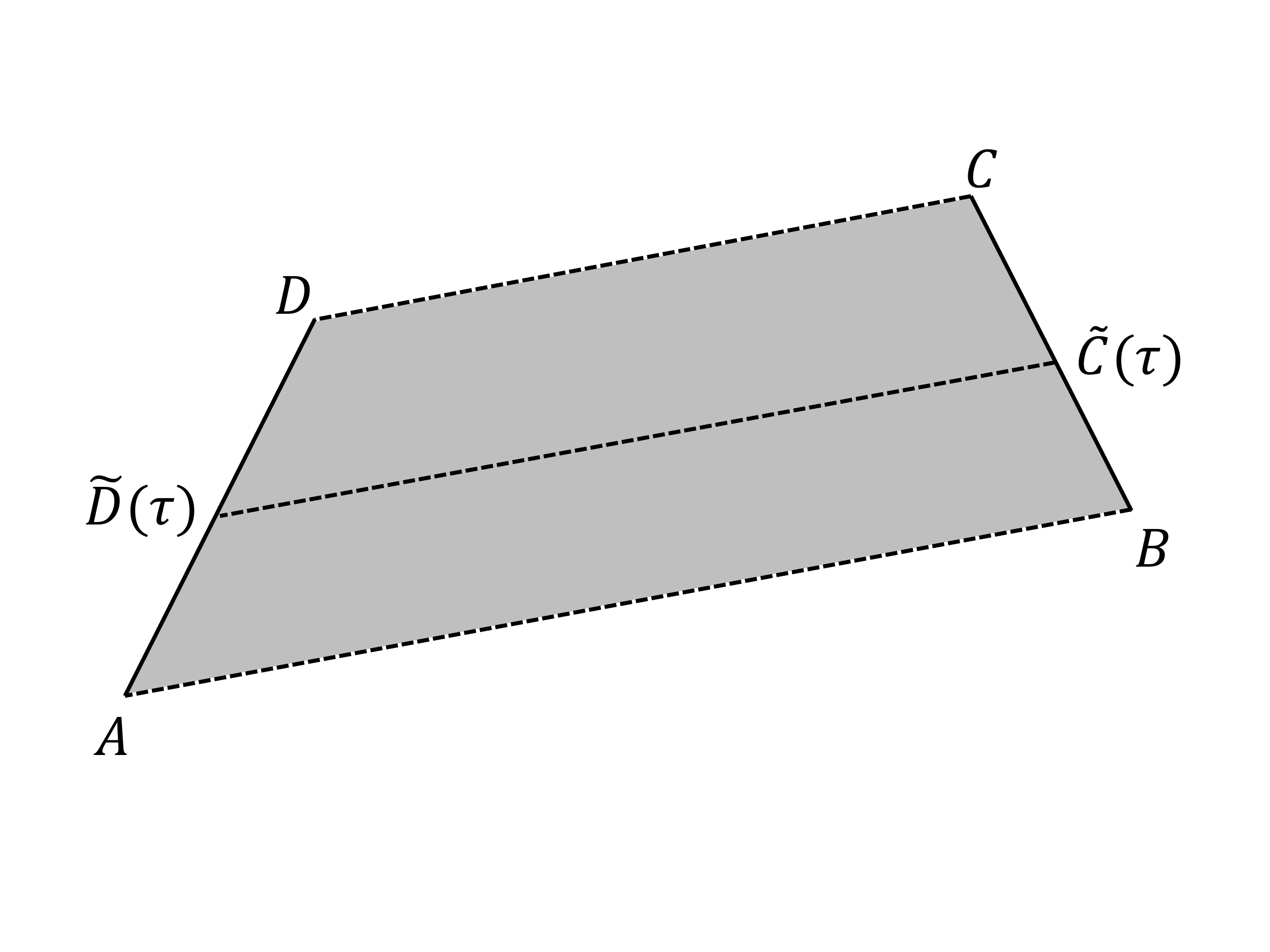}
		\caption{}
		\label{subfig:quadrilateral}
    \end{subfigure}
	\caption{(a): Evolving face-interface intersection line (dashed) drawn for each time where it hits a vertex. An example of the area swept between two such times is marked (grey quadrilateral). (b): Auxiliary notation for calculation of face-interface intersection line at intermediate times.}
	\label{fig:fintersect}
\end{center}
\end{figure}

We now use $\mathbf x_S$, $\hat{\mathbf n}_S$, and $U_S$ to approximate the time evolution of the face-interface intersection line of a face $j$, which is \emph{downwind} of surface cell $i$. This we do by \org{finding}\new{estimating} the times at which the isoface, travelling with velocity $U_S$ normal to itself, will reach the vertex points of face $j$ (see Fig.~\ref{subfig:sweptArea}). 
Let us temporarily denote the $N$ vertex points of face $j$ by $\mathbf X_1,...,\mathbf X_N$, and the times at which the isoface passes these points by $t_1,...,t_N$. Then we can estimate these times as
\begin{equation}\label{eq:timeList}
	t_k \approx t + (\mathbf X_k-\mathbf x_S)\cdot \hat{\mathbf n}_S/U_S, \textrm{ for } k = 1,...,N.
\end{equation}
To obtain the face-interface intersection line at a given time $\tau\in[t,t+\Delta t]$, we can now apply a linear interpolation based edge cutting procedure equivalent to the one used to find the initial (i.e. at time $t$) isoface from the volume fractions. Only, now the function values at the vertices are the times from \eqref{eq:timeList}, rather than the interpolated volume fractions. 

More specifically, let us temporarily denote by $AB$ the line segment at time $t_k$, and by $CD$ the line segment at time $t_{k+1}$, such that $ABCD$ is the grey quadrilateral shown in Fig.~\ref{subfig:sweptArea} and \ref{subfig:quadrilateral}. Then at an intermediate time, $\tau\in[t_{k},t_{k+1}]$, we will assume the two end points of the \new{face-interface intersection line} segment to be
\begin{equation}\label{eq:DCtilde}
	\tilde D(\tau) = A + \frac{\tau - t_k}{t_{k+1}-t_k}(D-A) \textrm{ and } \tilde C(\tau) = B + \frac{\tau - t_k}{t_{k+1}-t_k}(C-B)
\end{equation}
as illustrated in Fig.~\ref{subfig:quadrilateral}. This concludes our approximation of the face-interface intersection line evolution during a time step.

\subsection{Time integral of submerged face area}\label{ssec:timeIntSubArea}
To calculate the time integral of the submerged area, $A_j(\tau)$ in \eqref{eq:DeltaVApprox2}, we first generate a sorted list of times, $\tilde t_1,...\tilde t_M$, starting with $\tilde t_1 = t$, and ending with $\tilde t_M = t+\Delta t$, and with all the $t_k$'s from \eqref{eq:timeList} satisfying $t<t_k<t+\Delta t$ in between. Then the time integral in \eqref{eq:DeltaVApprox2} can be split up as follows,
\begin{equation}\label{eq:timeIntASubTimes}
	\int_{t}^{t+\Delta t}A_j(\tau)d\tau = \sum_{k = 1}^M\int_{\tilde t_k}^{\tilde t_{k+1}}A_j(\tau)d\tau.
\end{equation}
On each of these subintervals, the face-interface intersection line sweeps a quadrilateral as the one shown in Fig.~\ref{subfig:quadrilateral}. Using the definition in \eqref{eq:DCtilde}, the submerged area at the intermediate time $\tilde t_k\leq\tau\leq \tilde t_{k+1}$ is
\begin{eqnarray}
	A_j(\tau) & = & A_j(\tilde t_k) + \frac12\textrm{sign}(U_S)\left|A\tilde C(\tau)\times B\tilde D(\tau)\right| \nonumber \\
	& = & P_k\tau^2+Q_k\tau+A_j(\tilde t_k).
\end{eqnarray}
Here $P_k$ and $Q_k$ are polynomial coefficients that can be calculated analytically from $A, B, \tilde C$ and $\tilde D$. The sign of $U_S$ in cell $i$ accounts for the direction \new{of} propagation of the isoface, i.e. whether the cell and face are gaining or loosing fluid A during the time interval. Once these coefficients are obtained, the contribution to the time integral in \eqref{eq:timeIntASubTimes} from the sub time interval $[\tilde t_k,\tilde t_{k+1}]$ is simply
\begin{equation}
	\int_{\tilde t_k}^{\tilde t_{k+1}}A_j(\tau)d\tau = \frac13 [\tilde t_{k+1}^3-\tilde t_k^3] P_k+\frac12[\tilde t_{k+1}^2-\tilde t_k^2]Q_k+[\tilde t_{k+1}-\tilde t_k]A_j(\tilde t_k)
\end{equation}
Adding up all these sub interval contributions, as devised by \eqref{eq:timeIntASubTimes}, and substituting the result into \eqref{eq:DeltaVApprox2}, we finally reach the sought estimate for $\Delta V_j(t,\Delta t)$.

As stated in Section \ref{ssec:algorithm}, the above procedure should be repeated for all downwind faces of a surface cell. On all other faces populate $\Delta V_j$ with the volume fraction of their upwind cell. The updated $\alpha_i$'s at time $t+\Delta t$ can now be calculated by inserting the $\Delta V_j$'s into \eqref{eq:finalAlpha}.

\subsection{Bounding procedure}\label{ssec:bounding}
The procedure described above gives an accurate estimate of the fluid transport across faces in many simple cases. It does, however, not guarantee strict boundedness, that is, there is nothing preventing the algorithm from producing updated volume fractions outside the physically meaningful range $0\leq\alpha_i(t+\Delta t)\leq 1$. Experience shows that slight unboundedness may be produced in cells just behind (i.e. upwind of) the interface. The explanation for this is that, while the method's estimate of the $dV_j$'s are typically very good, there will inevitably be small errors, which, in cases where a cell is completely emptied or filled during the time step, will cause the algorithm to miss $0$ or $1$ by a small amount. If the produced over- and undershoots are sufficiently small, one might be tempted to simply introduce a step in the algorithm that chops $\alpha_i(t+\Delta t)$ at $0$ and $1$ before proceeding to the next time step. However, since this corresponds to removing and adding fluid in cells, this method destroys strict volume conservation and is not true to the VOF idea of only allowing \emph{redistribution} of fluid amongst cells. While such a step may be practically necessary in order to ensure strict boundedness, it should be used with caution as it may potentially cause severe lack of volume conservation, in particular for long duration simulations. Can we instead introduce a bounding procedure, which is not adding or removing fluid from the domain, but only redistributing it in order to achieve boundedness? In the following, we will first explain our upper bounding procedure for redistributing the surplus of fluid A in cells with $\alpha_i(t+\Delta t) > 1$. \new{Then we show how the exact same procedure can be used for lower bounding.}

\subsubsection{Upper bounding}
Cells with $\alpha_i(t+\Delta t) > 1$ are typically just upwind of the interface, in regions where the interface is moving into fluid B (i.e. $U_S > 0$). Therefore the cells just upwind of an overfilled cell $i$ are filled with fluid A, and are therefore not good candidates for taking over the surplus of fluid A in cell $i$ \new{in a redistribution step}. On the other hand, the cells just downwind of cell $i$ are only partially filled with fluid A, and are therefore able to receive cell $i$'s small surplus of fluid A. But if cell $i$ has more than one downwind cell, how should its surplus of fluid A be distributed among these? We argue as follows: The overshooting of cell $i$ starts at the time $t^* \in [t, t+\Delta t ]$, where the cell becomes filled, i.e. $\alpha_i(t^*) = 1$. From this time on, all its faces must be completely filled with fluid A. Therefore pure fluid A will flow through its downwind faces from time $t^*$ and onwards. It is therefore natural to pass cell $i$'s surplus of fluid A through its downwind faces using the face fluxes, $\phi_j$, as the weighting factors. So if the fluid A surplus in cell $i$ is $V^+$, and the cell has $N$ downwind faces with fluxes $\phi_1,... ,\phi_N$, then the fraction of $V^+$ pass\new{ed} on through the $j$'th of these \new{faces} should be $\phi_j /\sum_{k=1}^N \phi_k$. However, we will not permit more fluid A to be passed through face $j$ than $\phi_j(t) \Delta t$. Therefore, we will clip the extra flux through face $j$ to $\min(\phi_j\Delta t,V^+\phi_j /\sum_{k=1}^N \phi_k)$. If a face reaches its maximum fluid A transport capacity, so the surplus flux is clipped in this way, the result is that not all the surplus $V^+$ in cell $i$ is passed on to downwind cells in this first redistribution step. In that case the step is repeated to pass on the remaining surplus of fluid A through the remaining downwind faces, still using the $\phi_j$'s as weightings, and clipping if the maximum capacity of a face is reached. The step is repeated until either all surplus fluid A in cell $i$ is passed on to the downwind neighbours, or there are no more downwind cells that can take up more fluid A. 

\subsubsection{Lower bounding}
The procedure for lower bounding (i.e. correcting cells with $\alpha_i(t+\Delta t) < 0$) follows simply by changing our perspective from that of fluid A to that of fluid B: We introduce the volume fraction of fluid B, $\beta_i \equiv 1-\alpha_i$, and the volume of fluid B transported across faces during $\Delta t$, $\Delta \tilde V_j \equiv \phi_j \Delta t - \Delta V_j$. Now $\alpha_i < 0$ is equivalent to $\beta_i > 1$ and we can apply the upper bounding procedure outlined above to correct the $\Delta \tilde V_j$'s. With the $\Delta \tilde V_j$'s corrected, we calculate $\Delta V_j = \phi_j \Delta t - \Delta \tilde V_j$ and insert in \eqref{eq:finalAlpha} to obtain the updated volume fraction $\alpha_i(t+\Delta t)$.

\subsubsection{Clipping}
It is our experience that the redistribution process outlined above succeeds in bounding most cells. However, occasionally all downwind faces of an overfilled cell will reach their maximum flux capacity before the cell is fully bounded. This only happens on rare occasions, and when it does it only has a minor effect on the overall quality of the solution. Nevertheless, some applications may require strict boundedness at all times, and so we have introduce an optional clipping of the volume fractions after the bounding procedure described above and before proceeding to the next time step. When this clipping is switched on the method is not strictly volume conserving, and one should therefore monitor the evolution of the total volume of fluid A, to ensure that it only varies within acceptable limits.

\section{Results}\label{sec:results}

In the following we present the results of simple test cases with isoAdvector. The numerically advected volume fractions should reproduce as accurately as possible, with the given mesh and time step size, the solution to an interface advection problem. A simple check of this is to advect a confined volume of fluid A across the computational mesh in a uniform velocity field, and observe to what extent the method preserves the shape of the volume as it should. 

The other type of test we will perform exploits the time reversibility of the advection problem: If we advect a confined volume of fluid A in a spatially and temporally varying velocity field for a period of time, the interface will be distorted. If we then reverse the flow, and run it backwards for the same amount of time, the volume should return to its initial position and shape. 

The following error measures will be used to quantify the solution quality:
\begin{itemize}
\item \emph{Shape preservation}. Our quantitative measure of shape preservation will be
\begin{equation}
	E_1(t) \equiv \frac{\sum_i V_i|\alpha_i(t)-\alpha_i^{\textrm{exact}}(t)|}{\sum_i V_i\alpha_i^{\textrm{exact}}(t)},
\end{equation}
where the $\alpha_i^{\textrm{exact}}$'s are the volume fraction representation of the known exact interface shape \new{at time $t$}.
\item \emph{Volume conservation}. The change in the total volume of fluid A in the domain relative to the initial fluid A volume,
\begin{equation}
	\delta V_{\textrm{rel}}(t) \equiv \frac{\sum_i \alpha_i(t)V_i-\sum_i \alpha_i(0))V_i}{\sum_i \alpha_i(0)V_i},
\end{equation}
should be zero in simulations, where no fluid A enters or leaves the domain.
\item \emph{Boundedness}. For the volume fractions to be physically meaningful, we should have \org{$0 \leq \min_i(\alpha_i)$ and $\max_i (\alpha_i )\leq 1$ at all times}\new{$0\leq \alpha_i\leq 1$ for $i = 1,...,N_C$. Our measures of unboundedness will be min$_i(\alpha_i)$ and max$_i(\alpha_i)$, where the minimum and maximum are taken over all cells at the end of a simulation.}
\item \emph{Sharpness}. For a sharp interface, the width of the region where $\alpha_i$ changes from $0$ to $1$ should be similar to the cell size. As quantitative sharpness measure, we use the volume between the $\alpha = 0.01$ and $0.99$ isosurfaces of the volume fraction data divided by the corresponding volume for the volume fraction representation of the exact solution. We will call this quantity $\delta W_{\textrm{rel}}$.
\item \emph{Efficiency}. Here we give the simulation times, $T_{\textrm{calc}}$\new{, in seconds}. All simulations were executed on \new{a single core of} an Intel Xeon 3.10GHz CPU (E5-2687W) on a Dell Precision T7600 Workstation.
\end{itemize}

\new{For benchmarking the isoAdvector algorithm, we compare its performance with three algebraic VOF schemes:
\begin{itemize}
\item Multidimensional Universal Limiter with Explicit Solution (MULES) \cite{deshpande_evaluating_2012}. This is the interface capturing method used in the OpenFOAM\textregistered{} interface flow solver, interFoam.
\item High Resolution Interface Capturing (HRIC) \cite{muzaferija1998two}. This scheme is for instance used in the commercial computational continuum mechanics software STAR-CCM+\textregistered.
\item Compressive Interface Capturing Scheme for Arbitrary Meshes (CICSAM) \cite{ubbink_method_1999}. This scheme is one of the available options in ANSYS Fluent\textregistered. In all subsequent CICSAM simulations, we use the recommended blending factor value $0.5$.
\end{itemize}
These schemes are chosen partly because of their wide use in practical engineering applications, and partly because they are developed for usage on arbitrary meshes. }

\new{All MULES calculations presented in the following were executed using the interFoam solver in OpenFOAM-2.2.0 with the velocity-pressure coupling calculation switched off. Since our CFD work is mainly based on OpenFOAM\textregistered{}, and our primary aim is to improve its interFoam solver, the main emphasis will be on benchmarking against MULES in the subsequent test cases.}

\new{For the HRIC and CICSAM calculations, we use our own implementations of the schemes in OpenFOAM\textregistered{}. The schemes are available together with isoAdvector code in the repository \cite{isoAdvector}, where all setup files for the following test cases may also be found.}

\subsection{Disk in steady uniform 2D flow}
\begin{figure}[!tb]
\begin{center}
\includegraphics[width=\linewidth]{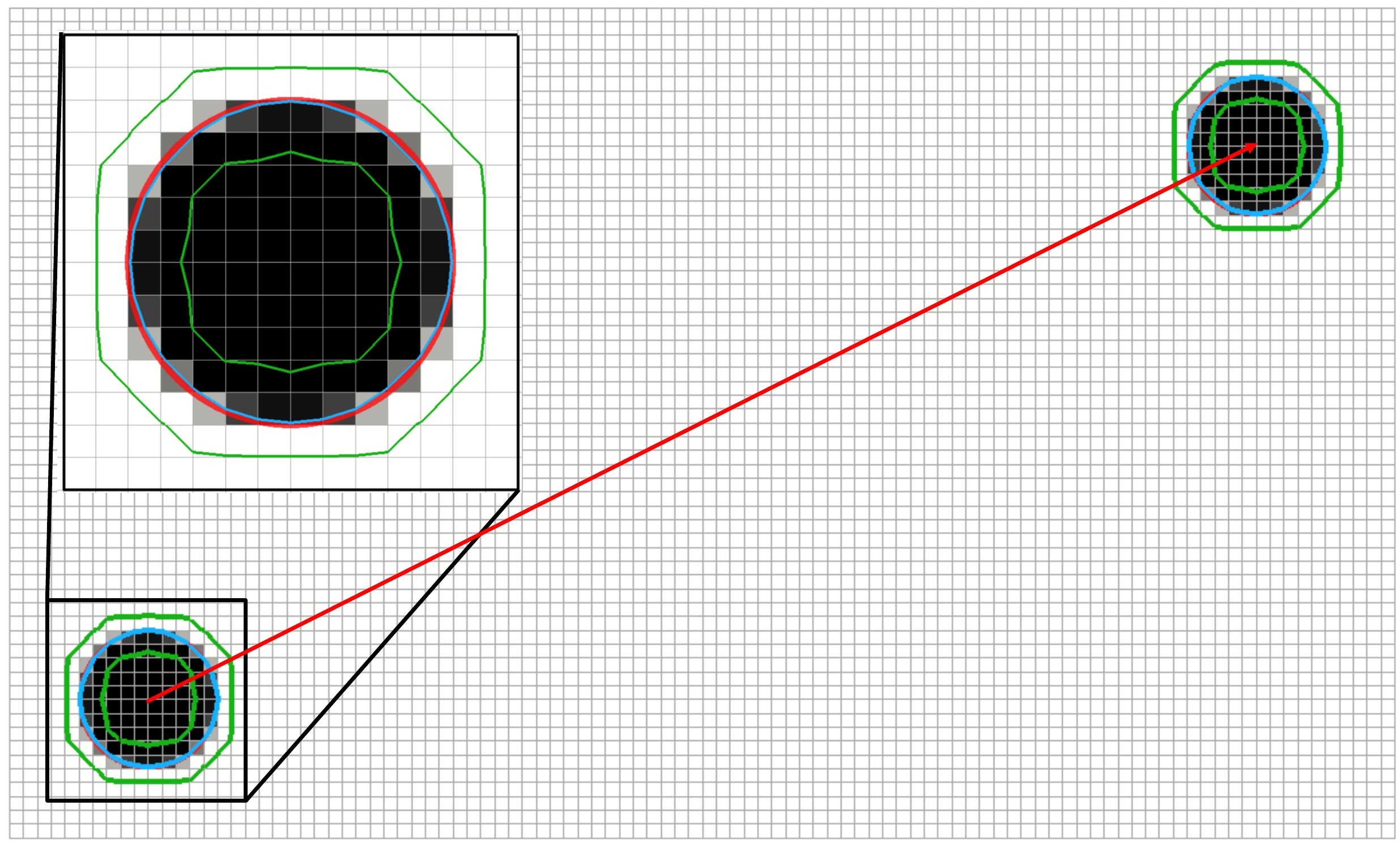}
\caption{A disk of fluid A of radius $0.25$ is initially centred at $(0.5, 0.5)$  (lower left corner). It moves with constant velocity $\mathbf u = (1, 0.5)$ for $4$ seconds ending at $(4.5, 2.5)$ (upper left corner). Volume fractions for the initial and final disk positions are shown with white cells being empty and filled cells begin black. Also shown are the $\alpha=0.5$ contour (blue) and the $\alpha=0.01$ and $0.99$ contours (green). A zoom of the initial condition is shown in upper left corner including a red circle marking the exact initial interface.}
\label{fig:discInUniFlowSetup}
\end{center}
\end{figure}

We start by considering a very simple 2D case on a mesh consisting of square cells: A circular region of fluid A of radius $R = 0.25$ moving in a constant and uniform velocity field, $\mathbf u = (1,0.5)$. The initial volume fractions are obtained from the R-isosurface of the function $\sqrt{(x-x_0)^2 + (y-y_0)^2}$, where $(x_0,y_0) = (0.5,0.5)$ is the initial position of the disk centre.
Fig.~\ref{fig:discInUniFlowSetup} shows the volume fraction representations of the exact initial and final interface \new{in grey scale, with white and black cells meaning empty and filled with fluid A, respectively}. 
The $\alpha = $ 0.5 contour is shown in blue, and the 0.01 and 0.99 contours are shown in green to indicate the minimal interface width on the given mesh resolution. In the top left corner of Fig.~\ref{fig:discInUniFlowSetup} we also show a zoom on the initial configuration with the exact circle shown in \org{blue}\new{red}. \org{For benchmarking the isoAdvector algorithm we here compare its performance with three other interface advection schemes:
\begin{itemize}
\item Multidimensional Universal Limiter with Explicit Solution (MULES) \cite{deshpande_evaluating_2012}. This is the interface capturing method used in the OpenFOAM\textregistered{} interface flow solver, interFoam.
\item High Resolution Interface Capturing (HRIC) \cite{muzaferija1998two}. This scheme is for instance used in the commercial computational continuum mechanics software STAR-CCM+\textregistered.
\item Compressive Interface Capturing Scheme for Arbitrary Meshes (CICSAM) \cite{ubbink_method_1999}. This is for instance on of the available schemes in ANSYS Fluent\textregistered.
\end{itemize}
These schemes are chosen partly because of their wide use in practical engineering applications and partly because they are developed for usage on general meshes.}

\subsubsection{Square meshes}
\begin{figure}[!b]
\begin{center}
\includegraphics[width=\textwidth]{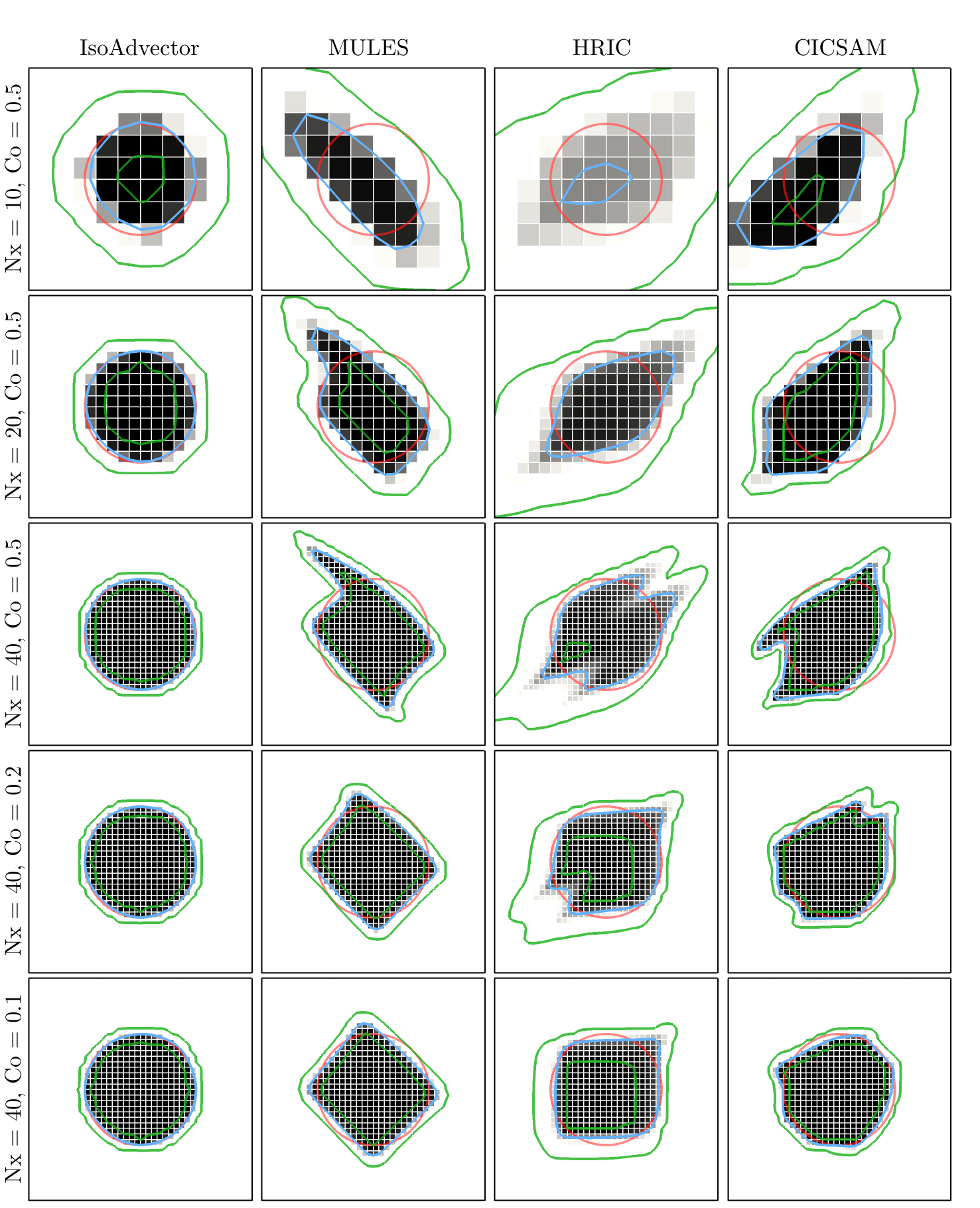}
    \caption{Disk in uniform flow $U = (1,0.5)$ at time $t = 4$ on a square mesh. Volume fractions shown in grey scale. Exact solution shown with red circles. $\alpha=0.5$ contour shown in blue, and $\alpha = 0.01$ and $\alpha = 0.99$ contours shown in green.}\label{fig:discInUniFlowHex}
\end{center}
\end{figure}

In Fig.~\ref{fig:discInUniFlowHex} we show in four columns (left to right) the final volume fraction solutions obtained with isoAdvector, MULES, HRIC, and CICSAM with 5 combinations of mesh and time resolution. In row 1-3 we investigate the effect of refining the mesh resolution with fixed Courant number, Co $= 0.5$. Then in row 3-5 we use the finest mesh and decrease Co from $0.5$ to $0.2$ and $0.1$. \new{Error measures and calculation times are displayed in Table~\ref{tab:discInUniFlow}.}
From Fig.~\ref{fig:discInUniFlowHex} and Table~\ref{tab:discInUniFlow} the following observations can be made:
\\ \\
\emph{Shape preservation}. The visual impression from Fig.~\ref{fig:discInUniFlowHex} is that isoAdvector is superior at preserving the shape of the disk on all shown mesh-Courant number combinations. MULES has a tendency to align the interface at 45 degree with the mesh faces. Therefore the MULES solution converges to a tilted square shape as cell and time step sizes are refined (2nd column in Fig.~\ref{fig:discInUniFlowHex}). The HRIC scheme shows a tendency to align the interface with the mesh faces, as also reported in \cite{nielsen_numerical_2003}. This causes the initially circular interface to converge to a square (3rd column in Fig.~\ref{fig:discInUniFlowHex}). For all the Co $= 0.5$ runs (4th column, row 1-3 in Fig.~\ref{fig:discInUniFlowHex}) CICSAM does not perform very well in terms of shape preservation. However, it is the only one of the reference schemes which converges to \new{something resembling} a circular interface solution as the time step is decreased (lower right corner in Fig.~\ref{fig:discInUniFlowHex}). Table~\ref{tab:E1_hex} quantifies these observations, showing that the isoAdvector $E_1$ error is at least a factor of $7$ smaller than the best of the other schemes for all runs. The table also reveals that the isoAdvector solution only improves slightly, when going from Co = 0.5 to Co = 0.2, and becomes slightly worse from Co = 0.2 to 0.1. Increasing errors with decreasing time step size \new{on a fixed mesh} was also reported in \cite{ubbink_method_1999}. \new{From the three Co = 0.5 errors in Table~\ref{tab:E1_hex}, we calculate isoAdvector's order of convergence with mesh refinement to be $\sim$ 2.4}. 
\\ \\
\FloatBarrier
\begin{table}[!p]
\begin{subtable}[htb]{\textwidth}
    \centering
        \begin{tabular}{|p{1.75cm}||p{1.75cm}|p{1.75cm}|p{1.75cm}|p{1.75cm}|}
            \hline
            (Nx,Co) & isoAdvector & MULES & HRIC & CICSAM \\
            \hline
            \hline
            (10,0.5) & 0.11 & 0.76 & 0.86 & 1 \\
            (20,0.5) & 0.035 & 0.43 & 0.56 & 0.73 \\
            (40,0.5) & 0.021 & 0.32 & 0.38 & 0.5 \\
            (40,0.2) & 0.014 & 0.2 & 0.22 & 0.27 \\
            (40,0.1) & 0.017 & 0.18 & 0.19 & 0.15 \\
            \hline
	      \end{tabular}
        \caption{$E_1$}
        \label{tab:E1_hex}
\end{subtable}
\begin{subtable}[htb]{\textwidth}
    \centering
        \begin{tabular}{|p{1.75cm}||p{1.75cm}|p{1.75cm}|p{1.75cm}|p{1.75cm}|}
            \hline
            (Nx,Co) & isoAdvector & MULES & HRIC & CICSAM \\
            \hline
            \hline
            (10,0.5) & 0 & -0.0019 & -0.026 & 0.0024 \\
            (20,0.5) & -6.7e-14 & -0.00018 & -0.001 & 0.0042 \\
            (40,0.5) & -3.6e-10 & -1.1e-05 & -1e-05 & 0.0012 \\
            (40,0.2) & -4.7e-13 & -5.2e-10 & 1.5e-05 & 0.00044 \\
            (40,0.1) & -2.1e-13 & -2.7e-13 & -8.8e-05 & 0.00019 \\
            \hline            \hline
	      \end{tabular}
        \caption{$\delta V_{\textrm{rel}}$}
        \label{tab:dVrel_hex}
\end{subtable}
\begin{subtable}[htb]{\textwidth}
    \centering
        \begin{tabular}{|p{1.75cm}||p{1.75cm}|p{1.75cm}|p{1.75cm}|p{1.75cm}|}
            \hline
            (Nx,Co) & isoAdvector & MULES & HRIC & CICSAM \\
            \hline
            \hline
            (10,0.5) & -3.9e-15 & 0 & 0 & -0.084 \\
            (20,0.5) & 0 & 0 & 0 & -0.24 \\
            (40,0.5) & 0 & 0 & -5.9e-13 & -0.23 \\
            (40,0.2) & 0 & 0 & 0 & -0.091 \\
            (40,0.1) & 0 & 0 & 0 & -0.047 \\
            \hline
	      \end{tabular}
        \caption{$\min_i(\alpha_i)$}
        \label{tab:amin_hex}
\end{subtable}
\begin{subtable}[htb]{\textwidth}
    \centering
        \begin{tabular}{|p{1.75cm}||p{1.75cm}|p{1.75cm}|p{1.75cm}|p{1.75cm}|}
            \hline
            (Nx,Co) & isoAdvector & MULES & HRIC & CICSAM \\
            \hline
            \hline
            (10,0.5) & 0 & 0.024 & 0.46 & -0.075 \\
            (20,0.5) & 0 & 3.4e-11 & 0.063 & -0.16 \\
            (40,0.5) & 0 & 0 & 0.0069 & -0.18 \\
            (40,0.2) & 0 & 4e-14 & 0.00035 & -0.1 \\
            (40,0.1) & 5e-14 & 6.4e-13 & 2.1e-05 & -0.034 \\
            \hline
	      \end{tabular}
        \caption{$1 - \max_i(\alpha_i)$}
        \label{tab:amax_hex}
\end{subtable}
\begin{subtable}[htb]{\textwidth}
    \centering
        \begin{tabular}{|p{1.75cm}||p{1.75cm}|p{1.75cm}|p{1.75cm}|p{1.75cm}|}
            \hline
            (Nx,Co) & isoAdvector & MULES & HRIC & CICSAM \\
            \hline
            \hline
            (10,0.5) & -0.0003 & 0.34 & 1.4 & 0.21 \\
            (20,0.5) & -0.0051 & 0.44 & 1.9 & 0.2 \\
            (40,0.5) & 0.026 & 0.51 & 3 & 0.27 \\
            (40,0.2) & 0.012 & 0.32 & 1.7 & 0.14 \\
            (40,0.1) & 0.0065 & 0.35 & 1.2 & 0.036 \\
            \hline
	      \end{tabular}
        \caption{$\delta W_{\textrm{rel}}$}
        \label{tab:dWrel_hex}
\end{subtable}
\begin{subtable}[htb]{\textwidth}
    \centering
        \begin{tabular}{|p{1.75cm}||p{1.75cm}|p{1.75cm}|p{1.75cm}|p{1.75cm}|}
            \hline
            (Nx,Co) & isoAdvector & MULES & HRIC & CICSAM \\
            \hline
            \hline
            (10,0.5) & 0.22 & 0.52 & 0.22 & 0.19 \\
            (20,0.5) & 0.85 & 2.16 & 0.94 & 0.87 \\
            (40,0.5) & 4.27 & 12.42 & 5.28 & 4.49 \\
            (40,0.2) & 7.41 & 28.28 & 9.46 & 8.4 \\
            (40,0.1) & 12.94 & 55.13 & 16.82 & 14.61 \\
            \hline
	      \end{tabular}
        \caption{$T_{\textrm{calc}}$}
        \label{tab:calcTimes_hex}
\end{subtable}
\caption{Performance for disk in uniform flow on a square mesh.}
\label{tab:discInUniFlow}
\end{table}
\FloatBarrier
\noindent \emph{Volume conservation}. From Table~\ref{tab:dVrel_hex}, we see that isoAdvector is the only scheme with volume preservation down to machine precision even on the coarsest mesh. On the finest mesh MULES also performs very good followed by HRIC, CICSAM being the worst performing scheme in this comparison. 
\\ \\
\emph{Boundedness}. From Table~\ref{tab:amin_hex} and \ref{tab:amax_hex}, we see that isoAdvector keeps the volume fraction data bounded to within machine precision. Also MULES and HRIC produce bounded volume fractions, whereas CICSAM has severe bounding problems even on the finest mesh.
\\ \\
\noindent\emph{Sharpness}. Table~\ref{tab:dWrel_hex} shows our sharpness measure, $\delta W_{\textrm{rel}}$. For all \org{runs}\new{simulations} the isoAdvector thickness is very close to the best one can expect, i.e. the thickness of the volume fraction representation of the exact solution on the given mesh. The MULES interface width is only 30-50\% larger than the width of the exact solution. HRIC performs rather bad in terms of interface sharpness with a smearing of the interface which is clearly visible in Fig.~\ref{fig:discInUniFlowHex} (column 3). CICSAM keeps the interface sharp for all runs and is the best performing of the reference schemes in this respect.
\begin{figure}[!b]
\centering
\includegraphics[width=\textwidth]{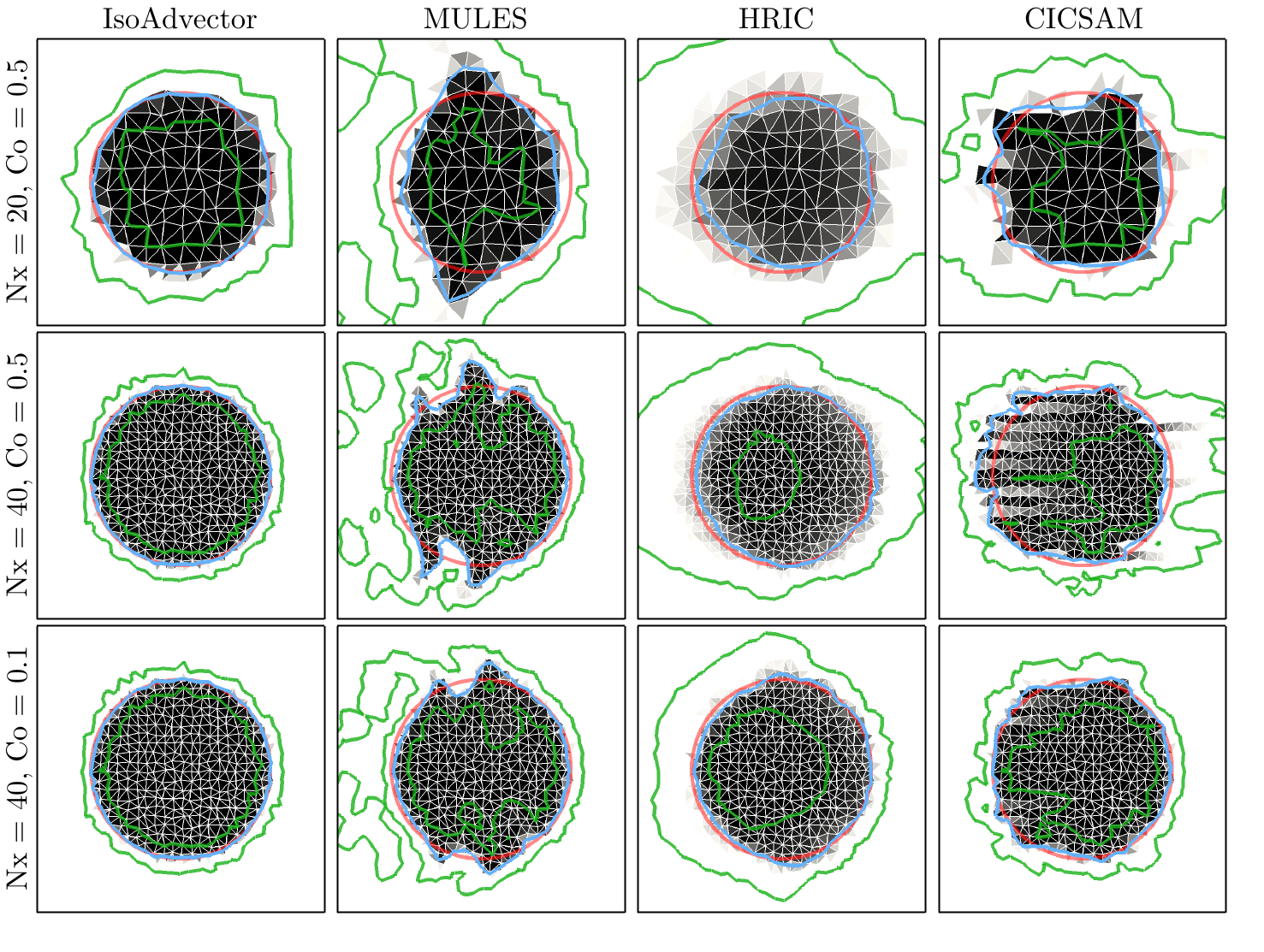}
    \caption{Disk in uniform flow $U = (1,0)$ at time $t = 4$ on a triangle mesh.}\label{fig:discInUniFlowTri}
        \begin{tabular}{|p{1.75cm}||p{1.75cm}|p{1.75cm}|p{1.75cm}|p{1.75cm}|}
            \hline
            (Nx,Co) & isoAdvector & MULES & HRIC & CICSAM \\
            \hline
            \hline
            (20,0.5) & 0.029 & 0.27 & 0.43 & 0.96 \\
            (40,0.5) & 0.014 & 0.18 & 0.26 & 0.26 \\
            (40,0.1) & 0.014 & 0.13 & 0.15 & 0.099 \\
            \hline
	      \end{tabular}
\captionof{table}{$E_1$ for simulations in Fig.~\ref{fig:discInUniFlowTri}. \new{Nx in the left column is the resolution of the square base mesh from which the unstructured meshes were generated by randomly distorting the points followed by a delaunay triangulation.}} \label{tab:discInUniFlowTri}
\end{figure}
\\ \\
\noindent\emph{Efficiency}. From Table~\ref{tab:calcTimes_hex}, we see that for this simple test case isoAdvector is \new{slightly faster than the fastest reference schemes, CICSAM and HRIC, for most simulations, and 2-4 times faster than MULES. It is remarkable, that the isoAdvector scheme can obtain a significantly improved accuracy with this significantly lower usage of computer resources than MULES.}\org{almost twice as slow as MULES and 4-6 times slower than HRIC and CICSAM. It is our experience that in interFoam, which uses MULES, typically 10-20\% of the calculation time is spent in the interface advection step. Doubling the time spend in this step would therefore correspondingly cause an increase in the overall calculation time of 10-20\%. Considering the gain in accuracy obtained with isoAdvector, this additional cost is considered to be acceptable for many applications.}
\subsubsection{Unstructured meshes}
\begin{figure}[!b]
\centering
\includegraphics[width=\textwidth]{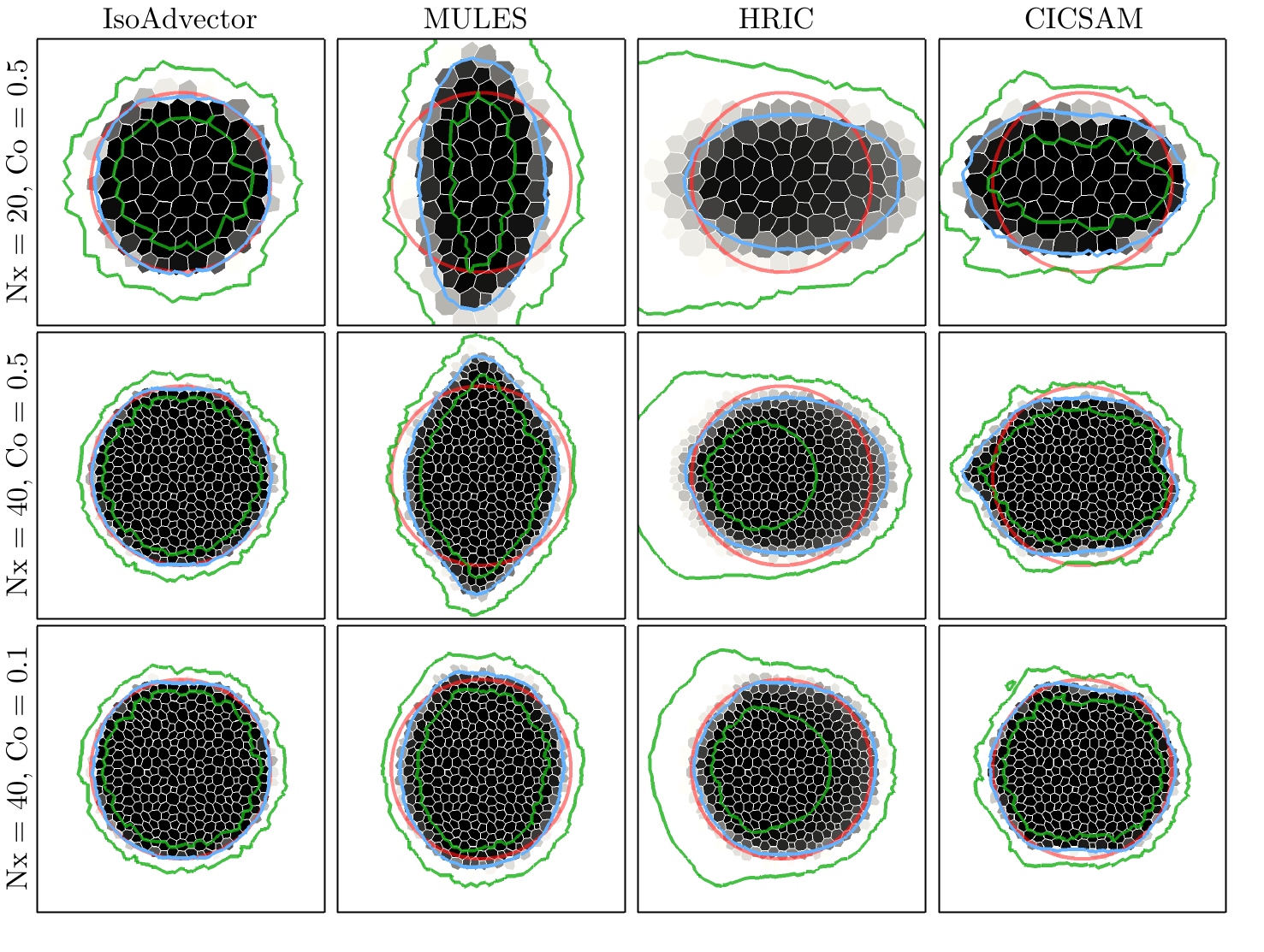}
    \caption{Disk in uniform flow $U = (1,0)$ at time $t = 4$ on a polygon mesh.}\label{fig:discInUniFlowPoly}
        \begin{tabular}{|p{1.75cm}||p{1.75cm}|p{1.75cm}|p{1.75cm}|p{1.75cm}|}
            \hline
            (Nx,Co) & isoAdvector & MULES & HRIC & CICSAM \\
            \hline
            \hline
            (20,0.5) & 0.039 & 0.43 & 0.45 & 13 \\
            (40,0.5) & 0.02 & 0.25 & 0.23 & 0.82 \\
            (40,0.1) & 0.024 & 0.13 & 0.14 & 0.26 \\
            \hline
	      \end{tabular}
\captionof{table}{$E_1$ for simulations in Fig.~\ref{fig:discInUniFlowPoly}. \new{Meshes are dual meshes of triangle meshes in Fig.~\ref{fig:discInUniFlowTri}}} \label{tab:discInUniFlowPoly}
\end{figure}
In Fig.~\ref{fig:discInUniFlowTri} and \ref{fig:discInUniFlowPoly}, we show a sequence of simulations similar to those in Fig.~\ref{fig:discInUniFlowHex}, but now on triangular and polygonal meshes, respectively. Again the columns show (from left to right) the solutions obtained with isoAdvector, MULES, HRIC and CICSAM. From row 1 to 2, we refined the mesh, keeping the Courant number at $= 0.5$. From row 2 to 3, we \org{keep}\new{retain} the mesh, but go from Co = 0.5 to 0.1. Since the meshes have no preferred direction, we use velocity $\mathbf u = (1,0)$ for these simulations. The disk radius is still $R = 0.25$, and the solutions are shown at time $t = 4$. Inspection of Fig.~\ref{fig:discInUniFlowTri} and \ref{fig:discInUniFlowPoly} and the quantitative measures (here only $E_1$ is shown in Tables \ref{tab:discInUniFlowTri} and \ref{tab:discInUniFlowPoly}) reveals that most of the observations listed above for the square mesh also hold for the triangle and polygon meshes. \org{Some exceptions are}\new{There are, however, a number of differences concerning the performance of the reference schemes}:
\\ \\
First, the tendency of MULES to align the interface at $45$ degree with the mesh faces is no longer visible due to the random face orientations, which presumably causes this systematic error to cancel out. However, on the triangle mesh MULES still does not seem to converge to a circular interface due to the development of ``wings'' on the sides (relative to the flow direction) of the fluid A region. On the polygon mesh, MULES does significantly better in terms of shape preservation, though with a tendency to squeeze the interface along the direction of motion.
\\
\\
Second, HRIC is much better at preserving the interface shape on both the triangle and polygon mesh than on the square mesh. It is, however, still very diffusive. \new{This is a good example of a case, where the $\alpha=0.5$ contour (blue) alone would give a good impression of the performance, but where the $\alpha = 0.01$ and $0.99$ contours (green) reveal the excessive smearing of the interface.}
\\
\\
Third, CICSAM performs very poorly on the triangle mesh with threads of fluid B piercing into the disk volume from behind. On the polygon mesh, these threads are not present and the solution quality is similar to the square mesh solution. On both the triangle and the polygon mesh, CICSAM has the same problems with unboundedness that we saw on the square mesh.
\\
\\
\new{We conclude, that also on unstructured meshes in 2D the performance of isoAdvector is significantly better than the reference schemes with calculation times that are similar to HRIC and CICSAM, and significantly lower than MULES.}
\begin{figure}[!b]
\begin{center}
		\includegraphics[width=0.5\linewidth]{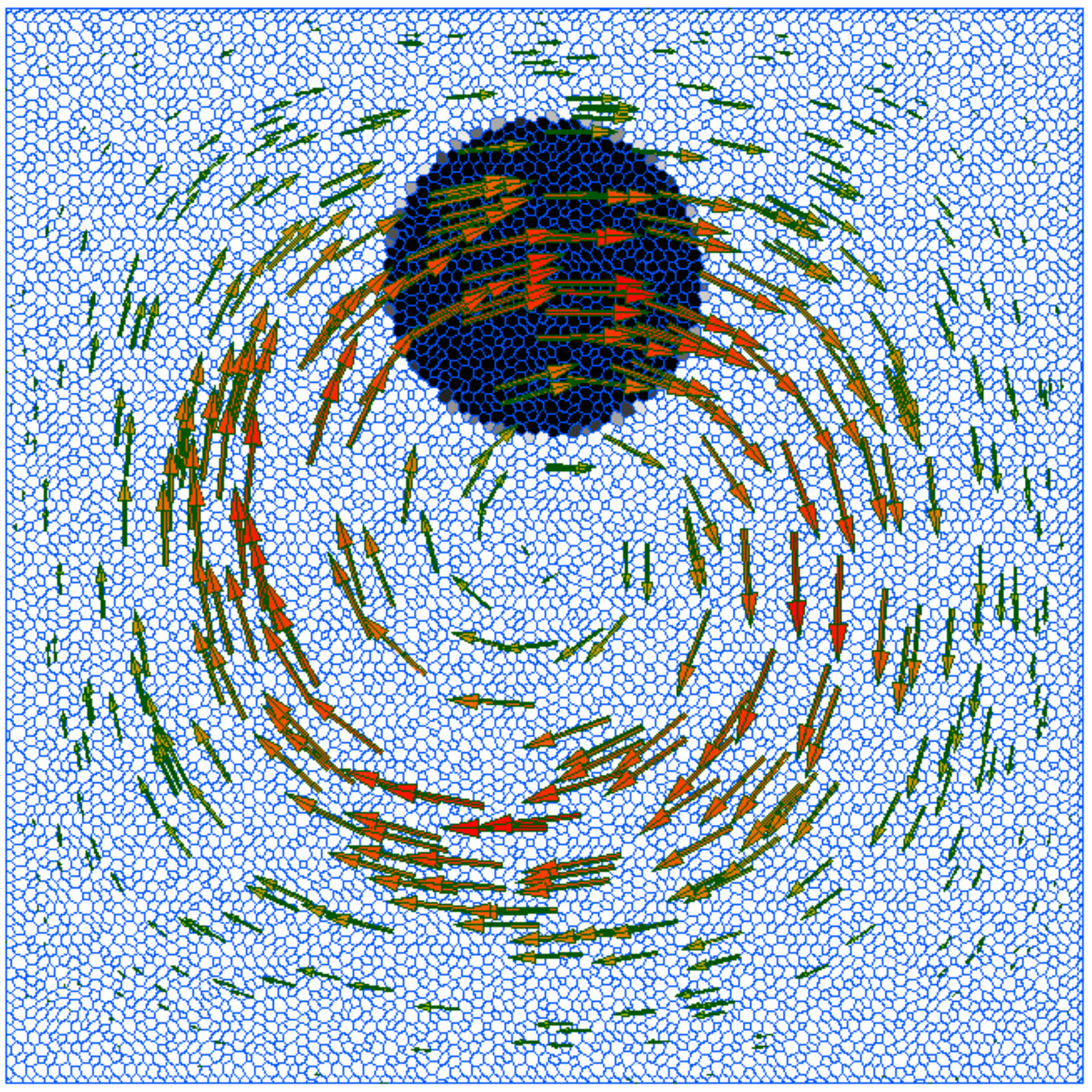}
        \caption{Initial condition for spiralling disk test case. }
        \label{fig:spiralDiscSetup}
\end{center}
\end{figure}
\subsection{Spiralling disc}
After 2D uniform flow tests, our next step is to test the solver performance in a spatially varying flow. We adopt the setup shown in Fig.~\ref{fig:spiralDiscSetup}, which has become a standard case for testing the ability of an interface advection schemes to deal with severe interface stretching\cite{ubbink_method_1999, jemison_coupled_2013, ahn_adaptive_2009, le_chenadec_3d_2013, harvie_new_2000, rider_reconstructing_1998, rudman_volume-tracking_1998, rudman_volume-tracking_1997}. The domain is the unit square with a disk of radius $R = 0.15$ initially placed at $(x,y) = (0.5,0.75)$. The velocity field is given by
\begin{equation}
	\mathbf u(x,y,t) = \cos(2\pi t/T)\left(-\sin^2(\pi x)\sin(2\pi y),\ 
	\sin(2\pi x)\sin^2(\pi y)\right),
\end{equation}
where the period of the flow is set to $T = 16$. This flow stretches the disk into a long filament until, at time $t = 4$, the flow is completely attenuated by the temporally varying cosine prefactor. Then the flow reverses, and the volume of fluid flows back into its original shape at time $t = 8$. At this time our shape preservation error measure, $E_1$, can be calculated by comparing the computed final state with the initial state. \new{Since we know the flow in advance, we use the fixed intermediate velocity, $\mathbf u(x,y,t+0.5\Delta t)$, on the whole time interval $[t,t+\Delta t]$.}

\begin{figure}[!b]
\begin{center}
	\includegraphics[width=.8\linewidth]{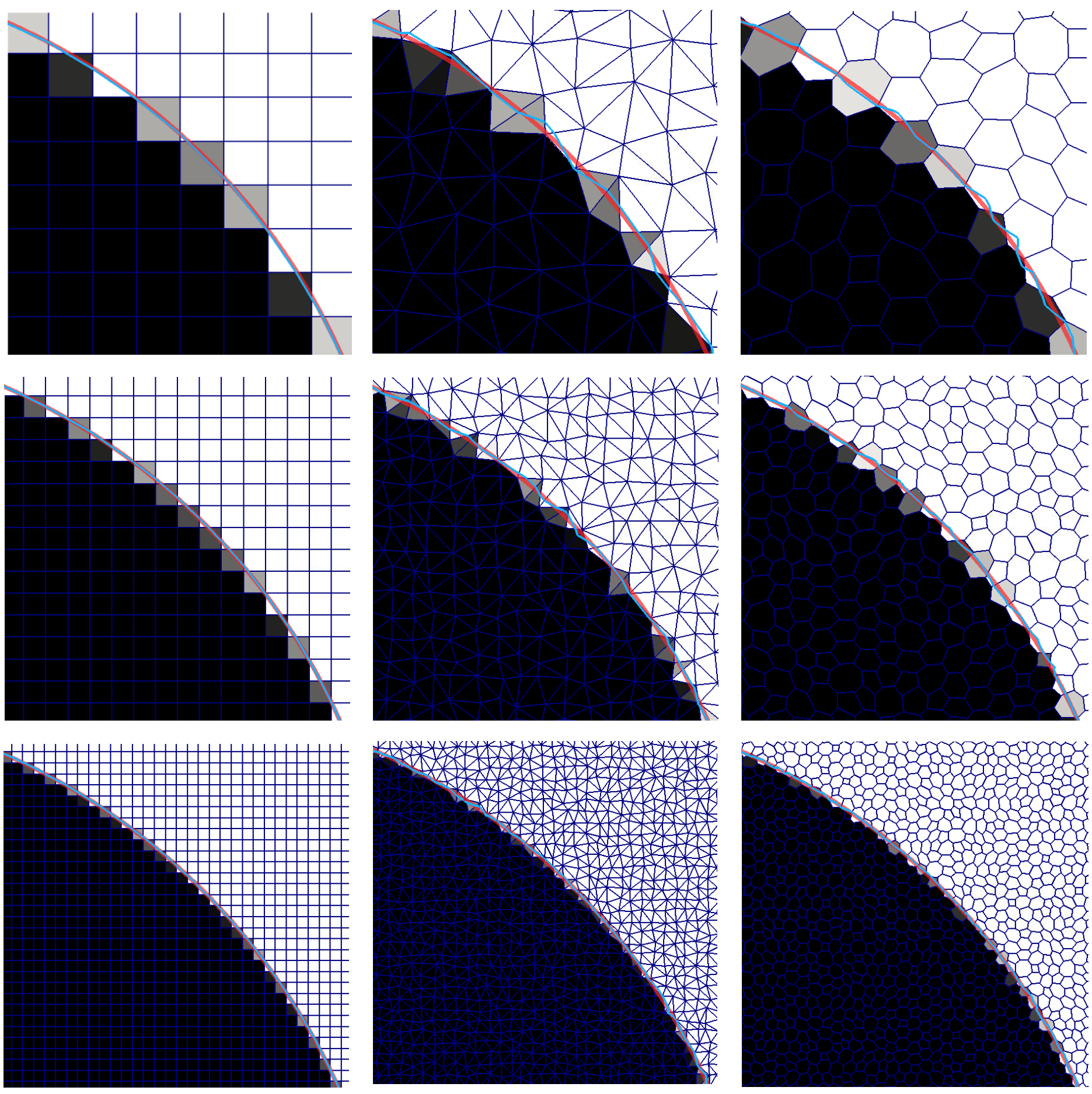}
	\caption{Meshes used to study spiralling disk case. Zoom on part of the initial interface. Exact circular shape shown in red and 0.5-contour of volume fractions shown in blue.}
        \label{fig:spiralDiscMeshes}
\end{center}
\end{figure}

\begin{figure}[!p]
\centering
\includegraphics[width=\linewidth]{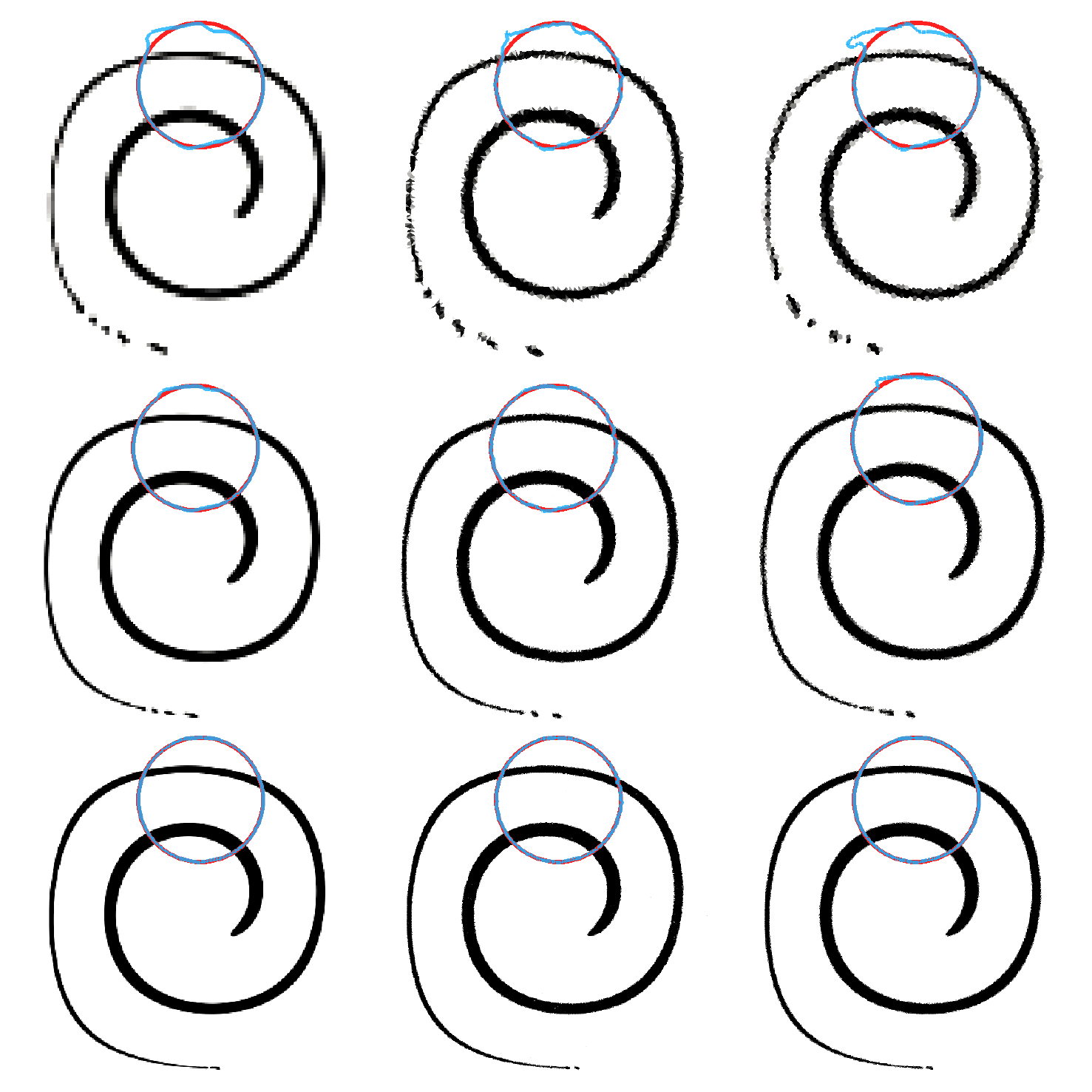}
\caption{Spiral disk isoAdvector results with Co = 0.5 on meshes from Fig.~\ref{fig:spiralDiscMeshes}.}
\label{fig:spiralDiscResults}
{\begin{subtable}{.49\textwidth}
\centering
\begin{tabular}{|p{.7cm}||p{1.1cm}|p{1.1cm}|p{1.1cm}|}
\hline
Mesh & square & triangle & polygon\\
\hline
\hline
1 & 10000 & 19602 & 10000\\
2 & 40000 & 79202 & 40000\\
3 & 160000 & 318402 & 160000\\
\hline
\end{tabular}
\caption{\#cells}
\label{tab:spirallingDiscIsoNcells}
\end{subtable}
\begin{subtable}{.49\textwidth}
\centering
    \begin{tabular}{|p{.7cm}||p{1.1cm}|p{1.1cm}|p{1.1cm}|}				        \hline
        Mesh & square & triangle & polygon\\
        \hline
        \hline
        1 & 0.047 & 0.054 & 0.071 \\
        2 & 0.012 & 0.02 & 0.018\\
        3 & 0.0023 & 0.0095 & 0.0039\\
        \hline
    \end{tabular}
    \caption{$E_1$}
    \label{tab:spirallingDiscIsoE1}
\end{subtable}
\begin{subtable}{.49\textwidth}
\centering
    \begin{tabular}{|p{.7cm}||p{1.1cm}|p{1.1cm}|p{1.1cm}|}
        \hline
        Mesh & square & triangle & polygon\\
        \hline
        \hline
        1 &  -1.5e-14 & -5.9e-15  &  -5e-15 \\
        2 &  3.4e-14 & -9.6e-14 & -1.8e-14\\
        3 & 2.1e-14 & -4.6e-13 & -1.8e-13\\
        \hline
    \end{tabular}
    \caption{$\delta V_\textrm{rel}$}
    \label{tab:spirallingDiscIsodV}
\end{subtable}
\begin{subtable}{.49\textwidth}
\centering
    \begin{tabular}{|p{.7cm}||p{1.1cm}|p{1.1cm}|p{1.1cm}|}
        \hline
        Mesh & square & triangle & polygon\\
        \hline
        \hline
        1 &  -6.1e-08 & -7.2e-09  & 0 \\                                   
        2 &  -2.8e-07 & -1.6e-10 & 0\\
        3 &  -4.7e-07 & -8.4e-12 & 0\\
        \hline
    \end{tabular}
    \caption{min$_i(\alpha_i)$}
    \label{tab:spirallingDiscIsoMinA}
\end{subtable}
\begin{subtable}{.49\textwidth}
\centering
    \begin{tabular}{|p{.7cm}||p{1.1cm}|p{1.1cm}|p{1.1cm}|}
        \hline
        Mesh & square & triangle & polygon\\
        \hline
        \hline
        1 &  -5.1e-08 & 0 & 0 \\                             
        2 &  -1.8e-08 & 0 & 0 \\                             
        3 &  -1.4e-08 & 0 & 0 \\
        \hline
    \end{tabular}
    \caption{1-max$_i(\alpha_i)$}
    \label{tab:spirallingDiscIsoMaxA}
\end{subtable}
\begin{subtable}{.49\textwidth}
\centering
    \begin{tabular}{|p{.7cm}||p{1.1cm}|p{1.1cm}|p{1.1cm}|}
        \hline
        Mesh & square & triangle & polygon\\
        \hline
        \hline
        1 & 13 & 58 &  26\\
        2 &  60 & 304 & 133\\
        3 & 314 & 1815 & 718\\
        \hline
    \end{tabular}
    \caption{$T_\textrm{calc}$}
    \label{tab:spirallingDiscIsoTcalc}
\end{subtable}
}\captionof{table}{Spiralling disk isoAdvector simulations with Co = 0.5.}
\label{tab:spirallingDiscIso}
\end{figure}

In Fig.~\ref{fig:spiralDiscMeshes} we show the square, triangle and polygon meshes in three different resolutions on which the isoAdvector method was tested. The results are shown in Fig.~\ref{fig:spiralDiscResults} using the same arrangement of the meshes. All simulations are run with Co = 0.5. In each panel, the exact initial and final interface shape is shown with a red circle overlaid with the $\alpha = 0.5$ contour (blue) of the final (i.e. at time $t = 8$) volume fraction data. The spiral shaped volume of fluid at time $t = 4$, where it is maximally stretched, is also shown in each panel. 
\setcounter{figure}{9}
All simulations show some degree of pinching at $t=4$. This occurs when the filament thickness reaches the cell size as is to be expected. The phenomena is therefore most pronounced on the coarsest meshes. We note that whereas the exact mathematical solution does not pinch, the 0.5-contour of its volume fraction representation will indeed pinch, if the mesh is coarse enough. As such, pinching does not have to be an error. However, as droplets pinch off, and the local interface curvature becomes comparable to the cell size, the isofaces are not able to represent the significant interface curvature inside a cell. The isoface based approximation of the advection then becomes faulty, leading to errors in the estimate of the droplet motion similar to those reported in \cite{cerne_numerical_2002}. The irreversibility of the introduced errors causes a distortion of the final disk in its upper region, which is made up of the previously pinched--off fluid. 

The mesh sizes, error measures and calculation times are shown in Table~\ref{tab:spirallingDiscIso}. \org{Roughly speaking the $E_1$ error is halved when the linear cell length is halved. The volume loss percentage vary from $10^{-7}$ down to machine precision. . Boundedness errors are zero to within machine precision. For the square and polygon meshes the interface width is only a few percent larger than the ``theoretical'' width, and $\delta W_\textrm{rel}$ is decreasing with increasing resolution. On the triangle mesh the interface width also decreases with increasing resolution but $\delta W_\textrm{rel}$ increases meaning that the ``theoretical'' width decreases faster. It is, however, evident from Fig.~\ref{fig:spiralDiscResults} that even on these very demanding triangle meshes isoAdvector yields very good results.}\new{From the $E_1$ values in Table~\ref{tab:spirallingDiscIso}, the orders of convergence with mesh refinement are calculated to be 1.9, 1.7 and 1.9 for the square, triangle and polygon meshes, respectively.}
\begin{figure}[!t]
\centering
		\includegraphics[width=\linewidth]{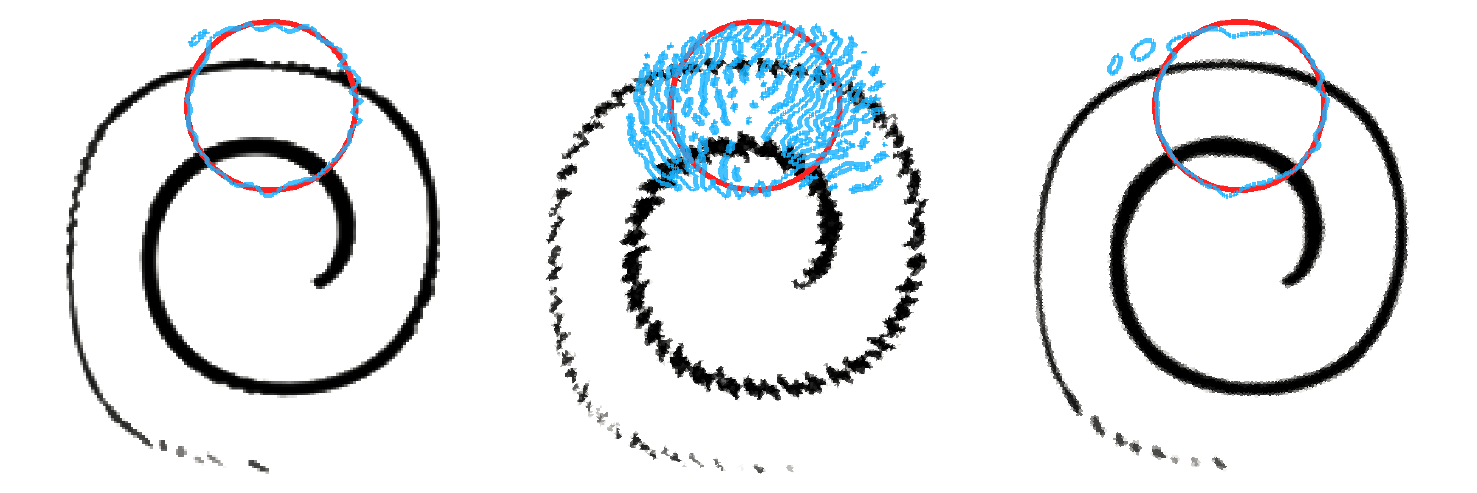}
        \caption{Spiral disk simulation MULES results with Co = 0.1 on the three intermediate resolution meshes from Fig.~\ref{fig:spiralDiscMeshes}.}
        \label{fig:spiralDiscMULES}
        \begin{tabular}{|p{1.75cm}||p{1.25cm}|p{1.25cm}|p{1.25cm}|p{1.7cm}|p{1.25cm}|p{1.25cm}|}
\hline
Mesh                        & square    & triangle  & polygon\\
\hline
\hline
$E_1$                       &  0.072    &  0.66        & 0.09\\
$\delta V_{\textrm{rel}}$   & 2.8e-14     & -3.1e-14  & -2e-15\\
$T_{\textrm{calc}}$         & 553       & 4151      & 1355 \\
\hline
	      \end{tabular}
        \captionof{table}{Spiralling disk error measures for MULES with Co = 0.1 on intermediate meshes (to compare with middle row in Fig.~\ref{fig:spiralDiscResults} and meshes 2 in Table~\ref{tab:spirallingDiscIso}).}
\label{tab:spiralDiscMULES}
\end{figure}
For a comparison, we show in Fig.~\ref{fig:spiralDiscMULES} and Table~\ref{tab:spiralDiscMULES} the results obtained with MULES on the intermediate resolution meshes of the three types, using Co = 0.1. For the square mesh, the MULES $E_1$ error is $\sim 50\%$ larger than the corresponding isoAdvector error. For the triangle mesh, the final interface is completely disintegrated. On the polygon mesh, MULES also gives acceptable results, although the $E_1$ error is 5 times larger than the isoAdvector error on the same mesh. In terms of calculation times, MULES is $\sim$10 times slower than isoAdvector. This is in part because MULES is run with smaller time steps. However, we also ran the simulations with Co = 0.5, in which case the MULES results on all three meshes were completely disintegrated like the triangle mesh solution in Fig.~\ref{fig:spiralDiscMULES}.

\subsection{Sphere in steady uniform 3D flow}
\begin{figure}[!t]
\begin{center}
	\includegraphics[width=.85\linewidth]{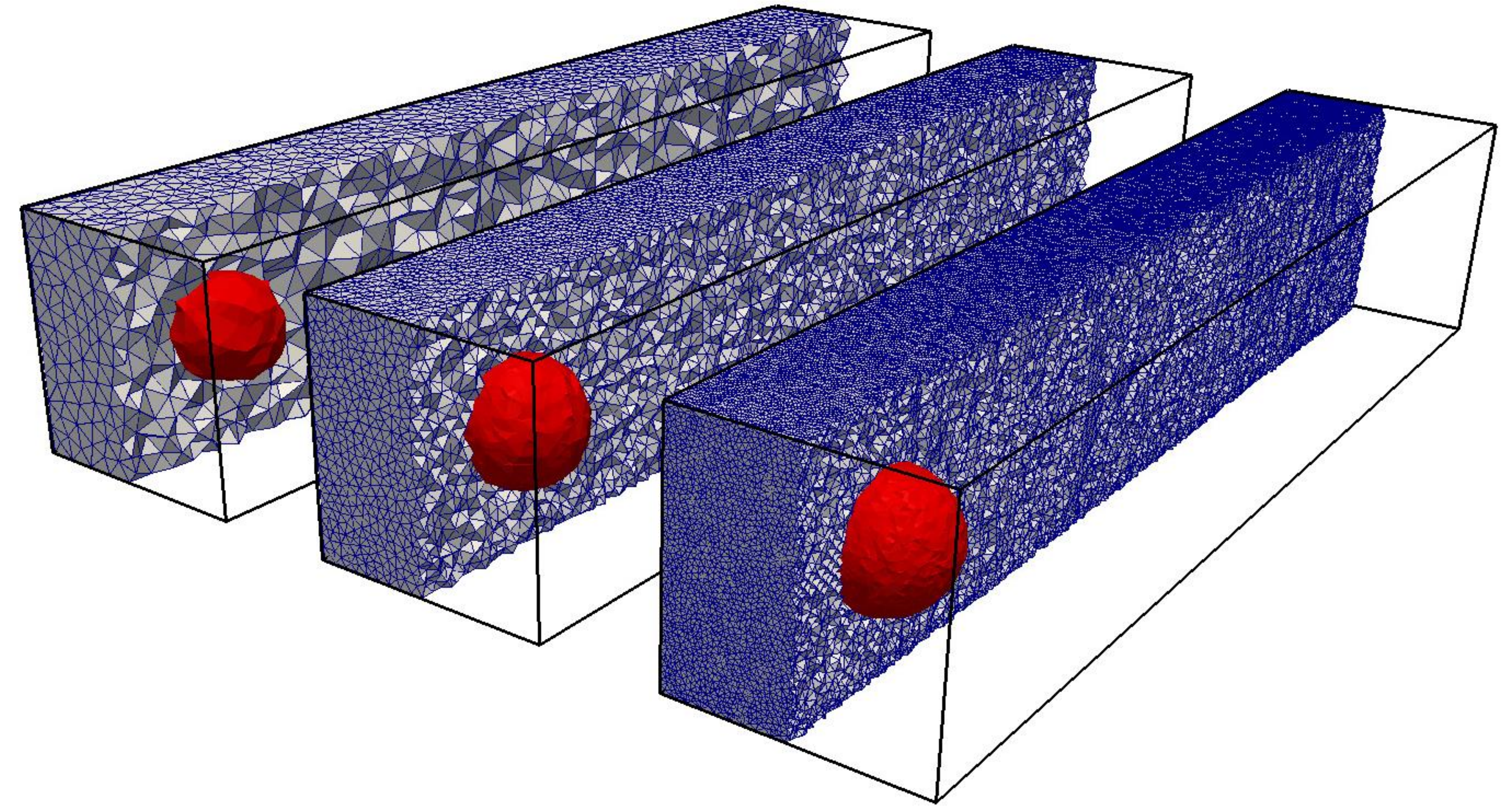}
    \caption{Random tetrahedron meshes used for sphere in uniform flow test case. $\alpha = 0.5$ isosurface shown for initial volume fraction data.}
    \label{fig:sphereInUniFlowTetMeshes}
\end{center}
\end{figure}
\begin{figure}[!h]
\begin{center}
	\includegraphics[width=.85\linewidth]{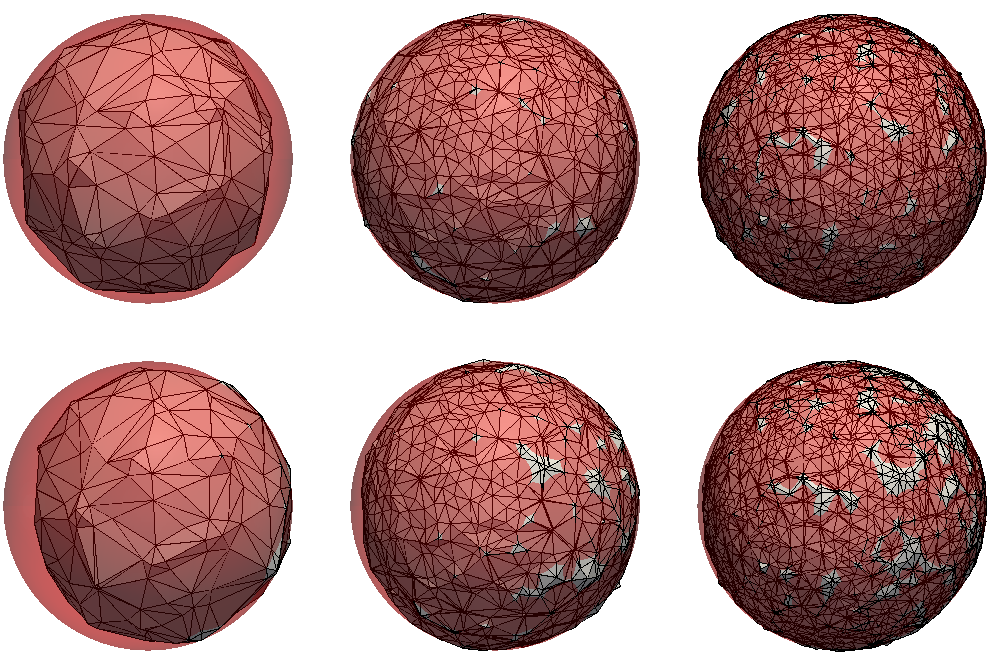}
    \caption{Sphere in uniform flow $U = (0,0,1)$ on tetrahedral mesh at time $t = 4$. Top row: Exact solution (red sphere) and its $0.5$-isosurface on three different mesh resolutions. Bottom row: Exact solution (red sphere) and $0.5$-isosurface of the isoAdvector solution with Co $= 0.5$.}\label{fig:sphereInUniFlowTet}
        \begin{tabular}{|p{1.25cm}||p{1.25cm}|p{1.25cm}|p{1.25cm}|p{1.7cm}|p{1.25cm}|p{1.25cm}|}
            \hline
            nCells & $E_1$ & $\delta V_{\textrm{rel}}$ & $\min_i(\alpha_i)$ & 1-$\max_i(\alpha_i)$ & $\delta W_{\textrm{rel}}$ & $T_{\textrm{calc}}$\\
            \hline
            \hline
            49.868 & 0.18 & -2.8e-11 & 0 & 0 & -0.033 & 11 \\
            343.441 & 0.046 & -7.8e-12 & -6.9e-11 & 0 & 0.0067 & 157 \\
            1.753.352 & 0.021 & 5.9e-11 & -2.7e-09 & 0 & 0.0035 & 1411 \\
            \hline
	      \end{tabular}
        \captionof{table}{Sphere in uniform flow errors and calculation times for isoAdvector with Co = 0.5.}
        \label{tab:sphereInUniFlowIsoOnTet}
\end{center}
\end{figure}
In this test we go back to a uniform flow, but now in 3D. The velocity is $U = (0,0,1)$, and the initial interface is a sphere of radius $R = 0.25$ centred at $(0.5,0.5,0.5)$. The simulations are run on three meshes consisting of 49.868, 343.441 and 1.753.352 random tetrahedra covering the domain, $[0,1]\times[0,1]\times[0,5]$. The meshes and the 0.5-isosurface of the initial volume fraction data are shown in Fig.~\ref{fig:sphereInUniFlowTetMeshes}. 
The simulations are run with Co = 0.5 until $t=4$, where the sphere has moved to $(0.5,0.5,4.5)$. The results are show in Fig.~\ref{fig:sphereInUniFlowTet} and in Table~\ref{tab:sphereInUniFlowIsoOnTet}. In the top row of Fig.~\ref{fig:sphereInUniFlowTet}, we show the exact final sphere (red) and the 0.5-isosurface of its volume fraction representation on the three mesh resolutions. In the bottom row, we show the exact sphere (red) together with the 0.5-isosurface of the final volume fraction data obtained with isoAdvector. As seen from Table~\ref{tab:sphereInUniFlowIsoOnTet}, the $E_1$ error on the coarsest mesh is fairly large. From Fig.~\ref{fig:sphereInUniFlowTet} (lower left panel), we see that this lack of overlap is mainly due to an overestimation of the propagation speed rather than a lack of ability to retain the spherical interface shape. On the finer meshes $E_1$ is reduced significantly, all though the tendency to be slightly ahead of the exact solution is still visible in Fig.~\ref{fig:sphereInUniFlowTet}. \new{The linear cell size is reduced by a factor 1.9 from the coarse to intermediate mesh, and by a factor 1.7 from the intermediate to fine. Based on these ratios, and the $E_1$'s in the Table~\ref{tab:sphereInUniFlowIsoOnTet}, the convergence order is in the calculated to be in the range 2.6-3.2.}

\begin{figure}[!h]
\begin{center}
	\includegraphics[width=.85\linewidth]{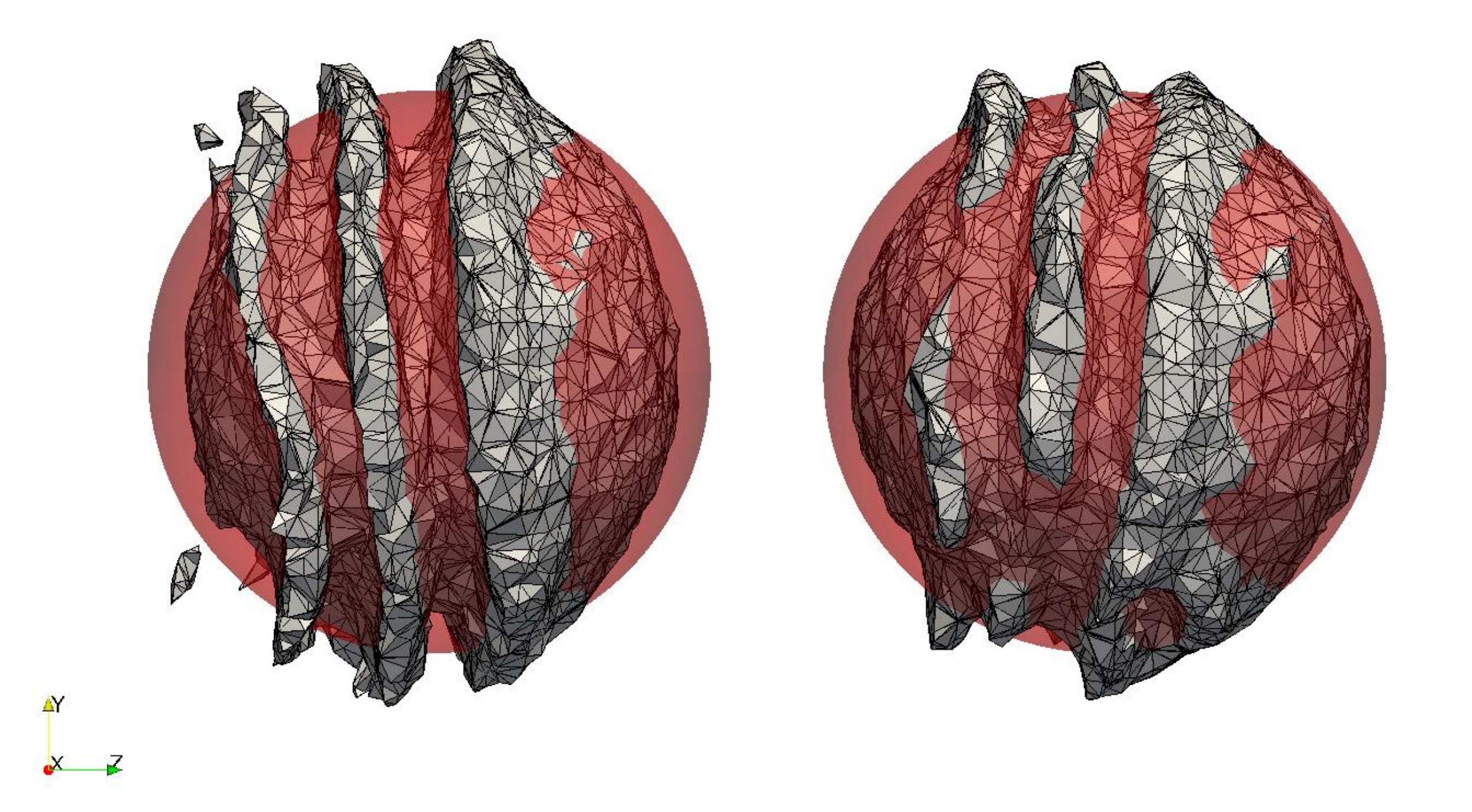}
    \caption{Sphere in uniform flow $U = (0,0,1)$ at time $t = 4$ on the finest tetrahedral mesh of Fig.~\ref{fig:sphereInUniFlowTet}. Left: Exact solution (red sphere) and $0.5$-isosurface of MULES solution with Co $= 0.5$. Right: The same but with Co $= 0.1$.}\label{fig:sphereInUniFlowTetMULES}
        \begin{tabular}{|p{1.25cm}||p{1.25cm}|p{1.25cm}|p{1.25cm}|p{1.7cm}|p{1.25cm}|p{1.25cm}|}
            \hline
            Co & $E_1$ & $\delta V_{\textrm{rel}}$ & $\min_i(\alpha_i)$ & 1-$\max_i(\alpha_i)$ & $\delta W_{\textrm{rel}}$ & $T_{\textrm{calc}}$\\
            \hline
            \hline
            0.5 & 0.42 & -4.5e-13 & -5e-06 & -1.9e-06 & 2 & 7306 \\
            0.1 & 0.29 & -8.2e-13 & 0 & 9e-06 & 1.9 & 28686 \\
            \hline
	      \end{tabular}
        \captionof{table}{Sphere in uniform flow errors measures for MULES on the finest tetrahedron mesh.}
        \label{tab:sphereInUniFlowMULESOnTet}
\end{center}
\end{figure}

For comparison, we show in Fig.~\ref{fig:sphereInUniFlowTetMULES} and Table~\ref{tab:sphereInUniFlowMULESOnTet} the results obtained with MULES on the finest mesh running with Co = 0.1 and 0.5. In both cases the shape preservation is significantly worse than the isoAdvector results. It is also noticeable that the MULES simulations with Co = 0.1 and Co = 0.5 are, respectively, 20 and 5 times slower than the corresponding isoAdvector simulation with Co = 0.5.
\subsection{Sphere in non-uniform 3D flow}
\begin{figure}[!t]
\begin{center}
\includegraphics[width=\textwidth]{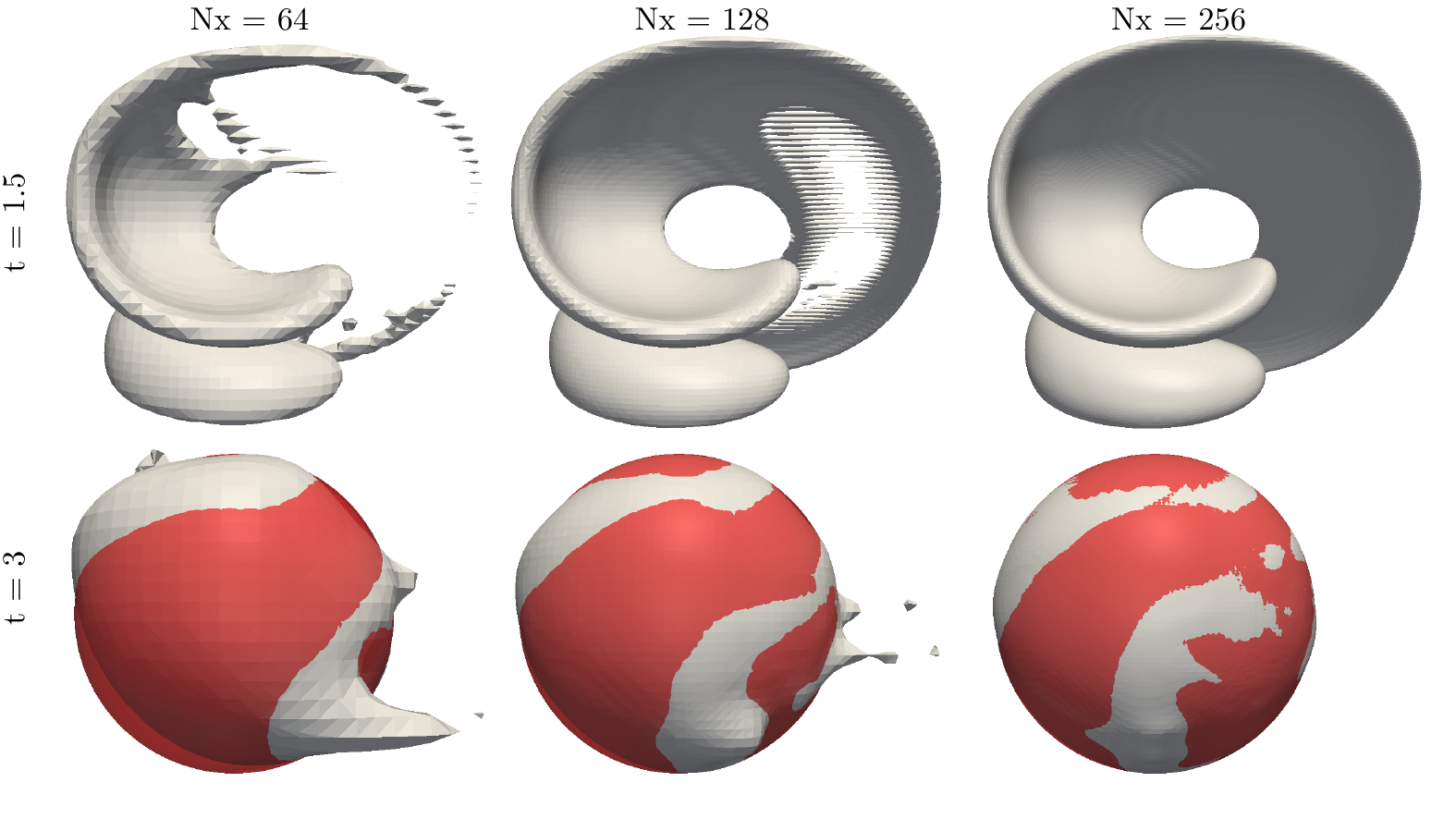}
    \caption{$\alpha = 0.5$ isosurfaces for sphere in non-uniform flow with Co = 0.5. Results for three mesh resolutions are shown at the time of maximum strechting, $t = 1.5$, and at the final time, $t = 3$. Exact final solution is shown with red spheres.}\label{fig:smearedSphere}
        \begin{tabular}{|p{1.25cm}||p{1.25cm}|p{1.25cm}|p{1.25cm}|p{1.7cm}|p{1.25cm}|p{1.25cm}|}
        	\hline
        	Nx & $E_1$ & $\delta V_{\textrm{rel}}$ & $\min_i(\alpha_i)$ & 1-$\max_i(\alpha_i)$ & $\delta W_{\textrm{rel}}$ & $T_{\textrm{calc}}$\\
        	\hline
        	\hline
        	$64$ & 0.22 & -2.6e-13 & 0 & 0 & 0.5 & 173\\
			$128$ & 0.047 & -2.6e-12 & -2.1e-11 & 0 & 0.17 & 2626\\
			$256$ & 0.012 & -1.6e-11 & -8.1e-09 & 0 & 0.12 & 46706\\
	        \hline
        \end{tabular}
        \captionof{table}{Error measures and calculation times for isoAdvector simulations in Fig.~\ref{fig:smearedSphere} of a sphere in a 3D non-uniform flow on three cube meshes.}
        \label{tab:smearedSphere}
\end{center}
\end{figure}
Our final test case is also in 3D, but now with a non-uniform velocity field. We adopt a setup often used to test surface smearing in 3D\cite{shin_local_2011,liovic_3d_2006,enright_fast_2005,jemison_coupled_2013}. The domain is the unit box, and the initial interface is a sphere of radius $R = 0.15$ centred at $(0.35,0.35,0.35)$. This surface is advected in the velocity field,
\begin{equation}\label{eq:3DvelocityField}
\mathbf u(x,y,z,t) = \cos(2\pi t/T)\left(
  \begin{array}{c}
  2\sin^2(\pi x)\sin(2\pi y)\sin(2\pi z) \\
  -\sin(2\pi x)\sin^2(\pi y)\sin(2\pi z) \\
  -\sin(2\pi x)\sin(2\pi y)\sin^2(\pi z)
  \end{array}
\right),
\end{equation}
where the period is set to $T = 6$. This flow stretches the sphere into a thin sheet creating two bending and spiralling ``tongues''. The maximum deformation is reached at $t = 1.5$, where the temporal cosine prefactor completely quenches the flow. From here on the flow reverses, and the interface returns to its initial shape and position at time $t = 3$. In Fig.~\ref{fig:smearedSphere} the isoAdvector results are shown at time $t = 1.5$ in the top row, and at time $t = 3$ in the bottom row, on three cubic meshes with $dx = 1/64, 1/128$ and $1/256$. In the lower panels, the exact final spherical shape is also shown in red. From ODE calculations with the velocity field \eqref{eq:3DvelocityField}, we have measured the sheet thickness at $t = 1.5$ to be $\sim0.0063$. This, and the fact that an edge can at most be cut once by the prescribed isosurface routine, explains why there are holes in the 0.5-isosurface of the volume fraction data on the two coarsest mesh with $dx \approx 0.016$ and $dx \approx 0.0078$, and no wholes in the finest simulation with $dx \approx 0.0039$. The error measures and calculation times for the three simulations are shown in Table~\ref{tab:smearedSphere}. \new{Based on the $E_1$'s in this table, the order of convergence is calculated to be 2.3.}
\begin{figure}[!t]
\begin{center}
	\includegraphics[width=\linewidth]{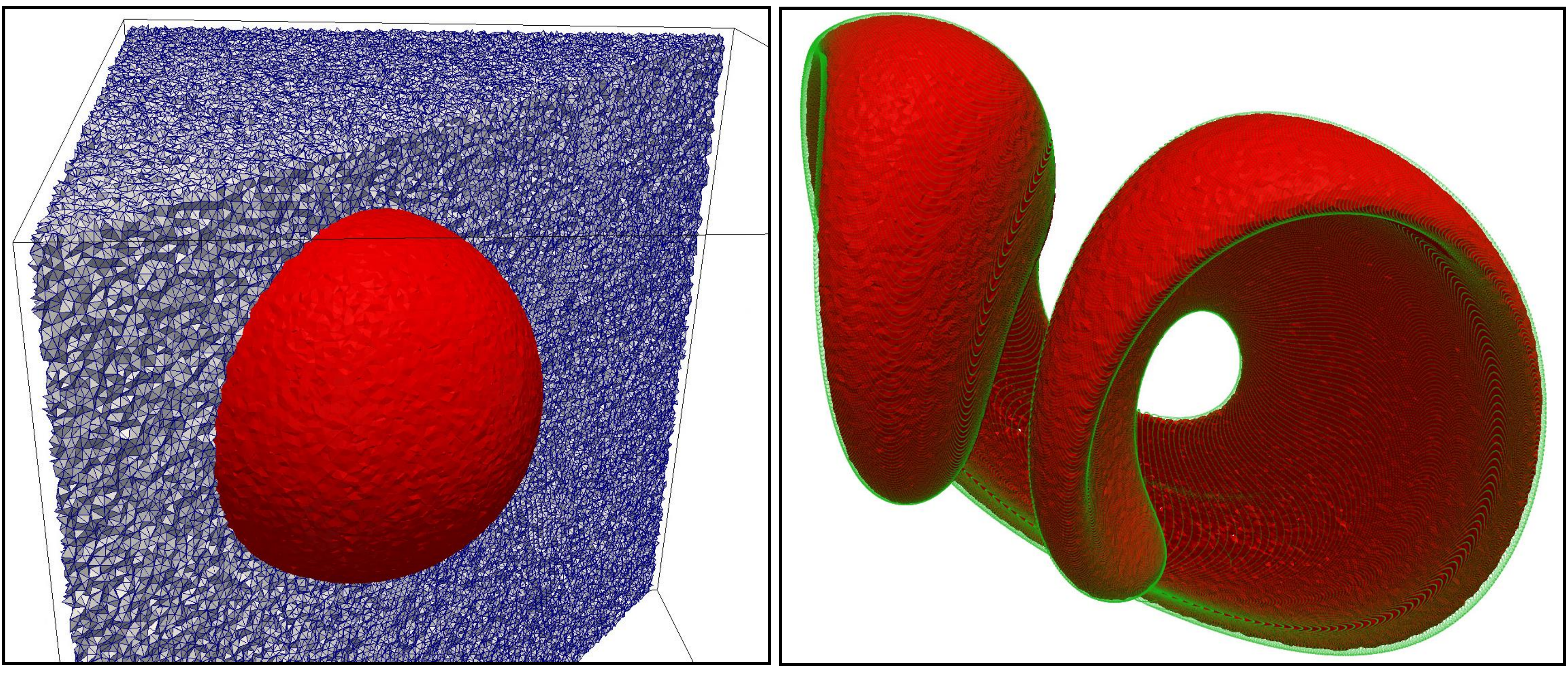}
    \caption{Sphere in non-uniform flow on tetrahedral mesh. Left: Mesh and 0.5-isosurface of the initial volume fraction data. Right: isoAdvector solution (red) and the solution obtained with an accurate ODE solver (green) at time $t = 1.5$.}\label{fig:smearedSphereTet}
\end{center}
\end{figure}
We have also performed this test on a mesh consisting of random tetrahedra. To get sufficient resolution to avoid holes in the 0.5-isosurface of the solution, we used a mesh with 10.131.041 cells. A cut through this mesh and the 0.5-isosurface of the volume fraction representation of the initial sphere are shown in the left panel on Fig.~\ref{fig:smearedSphereTet}. In the right panel, we show the isoAdvector solution at time $t = 1.5$, where the stretching is maximal. This panel also contains a solution obtained by integrating the velocity field with a Runge-Kutta ODE solver  for 160.000 points evenly distributed on the initial sphere (green dots). The visual impression from Fig.~\ref{fig:smearedSphereTet} is that there is a good match between the ODE and the isoAdvector solutions. Due to the clipping procedure in the bounding step, $\delta V_{\textrm{rel}}$ was 0.63\% at time $t = 3$. It took $\sim$3 days to simulate until $t = 3$, and it is therefore impractical to do further testing of isoAdvector on such large meshes until the code has been parallelised\org{ and profiled}. 

\section{Conclusion}\label{sec:conclusion}
We have developed a new algorithm, isoAdvector, for numerical interface advection across general \new{structured and unstructured} computational meshes. The method is derived from ``first principles'', i.e. from the control volume integrated continuity equation for a discontinuous density field. \new{The IsoAdvector scheme belongs to the class of geometric VOF methods, but with novel ideas implemented in both the interface reconstruction step, and in the interface advection step.}

\new{The novelty in the reconstruction step is the usage of efficient isosurface calculations to estimate the distribution of fluids inside computational cells. This is a very robust method even on unstructured meshes. It avoids the gradient calculations traditionally used in geometric VOF reconstruction step, which may cause problems, because the numerically estimated gradient is a cell volume averaged Dirac $\delta$-function. }

\new{In the interface advection step, the novelty is the division of the time step into sub time intervals on which we can analytically calculate the volume fraction flux through a mesh face under the assumption that the interface is moving steadily across the face during the interval. In the development of this procedure, no assumptions are made on the shape of a face, and therefore also the advection step is by design applicable on arbitrary meshes.} 

We have given a proof--of--concept by testing the method on various simple flow-interface combinations both in 2D and 3D structured and unstructured meshes. The results are very satisfactory both in terms of shape preservation, volume conservation, boundedness, interface sharpness, and efficiency. \new{The order of convergence with mesh refinement varies between 1.7 and 3.2 for the test case presented here}. Also, in spite of the geometric nature of some of the steps involved, the implementation of the new algorithm is relatively straightforward. 

\new{The isoAdvector advection step is explicit in nature, and the method is in principle limited to Courant numbers in the range $[0,1]$. In terms of accuracy, our experience so far indicates that the method has an optimum around Co $\sim$ 0.5 with only small degradation of solution quality when going to Courant numbers closer to 1. This is to be contrasted with the explicit MULES scheme in OpenFOAM\textregistered{}'s interFoam solver, which in our experience is limited to Co $\leq$ 0.1, if accuracy is important.\footnote{\new{It should be mentioned that from OpenFOAM\textregistered{} version 2.3.0 a new semi-implicit MULES scheme is introduced to solve some of the issues with boundedness, stability, and accuracy for large Courant numbers. The literature documenting the ideas behind and the performance of this new method is, however, still very sparse.}}}

The \new{isoAdvector} code\org{, isoAdvector--0.1\cite{isoAdvector},} is published\new{\cite{isoAdvector}} as an open source extension to OpenFOAM\textregistered{}. It is our hope that the isoAdvector concept and code will be used, tested, and further developed by the CFD community, and eventually result in improved simulation quality in the broad field of applications involving interfacial flows. 

\new{We note, that since the governing equation we solve is the passive advection equation for a scalar field in an solenoidal velocity field, the isoAdvector method may also find applications within other branches of CFD, where the advected surface is not necessarily marking the interface between two distinct fluids. There are many situations, where one needs to follow a passive tracer field, e.g. representing the concentration of some substance, which is immiscible with the surrounding fluid. Another possible application could be in an Immersed Boundary Method, where the isoAdvector scheme could provide accurate estimates of the fluid-solid interface within computational cells.}

\new{We are currently parallelising the isoAdvector code, and the parallelised version will be available in a new release in the code repository\cite{isoAdvector}. Based on the interFoam solver in OpenFOAM, we are also working on a consistent coupling of isoAdvector with a pressure-velocity solver. The performance of the resulting new interfacial flow solver will be presented in a future paper. Finally we note that, due to its applicability on arbitrary meshes, the isoAdvector code can be coupled with an adaptive mesh refinement routine with only minor modifications. Such a coupling will also be investigated in future work.}

\section*{Acknowledgment}

This work was sponsored by a Sapere Aude: DFF -- Research Talent grant from The Danish Council for Independent Research | Technology and Production Sciences to JR (Grant--ID: DFF -- 1337-00118). The grant also covers all activities of HB and HJ in connection with the project. JR also enjoys partial funding through the GTS grant to DHI from the Danish Agency for Science, Technology and Innovation. \new{We would like to express our sincere gratitude for this support.}\\
JR is grateful to the following people for fruitful discussions and for help on improving the isoAdvector code: Tomislav Mari\'{c}, Daniel Deising, and Holger Marschall from the Mathematical Modeling and Analysis Group at the Center of Smart Interfaces, Technische Universit\"at Darmstadt, Vuko Vuk\v{c}evi\'{c}$^3$, Tessa Uroi\'{c}$^3$, Bjarne Jensen$^1$, and Henrik Rusche from Wikki Ltd. 


\bibliographystyle{ieeetr}
\bibliography{isoAdvector}


\end{document}